\documentclass[11pt,draftclsnofoot,onecolumn]{IEEEtran}

\usepackage{fancyhdr}

\usepackage{ amsfonts}
\usepackage{ amsthm}
\usepackage{cite}

\usepackage{algorithmic,algorithm}

\usepackage{setspace} 

\usepackage[english]{babel}

\usepackage[tight]{subfigure}

\usepackage[cmex10]{amsmath}
\usepackage{amsfonts,amssymb}
\usepackage{bm}

\usepackage{array}
\usepackage{empheq}

\usepackage{dsfont}

\usepackage[dvips]{graphicx}

\usepackage{color}

\usepackage[letterpaper,top=1in,bottom=1in,left=0.75in,right=0.75in,bindingoffset=0in]{geometry}

\newcommand{\Reals}     {{{\mathrm I\!R}}}  

\newcommand{\define}    {\stackrel{\scriptscriptstyle\triangle}{=}}  

\newcommand{\uwti}[1]{{\mathbf #1}}

\newcommand{\cb}{{\uwti c}}

\newcommand{\hb}{{\uwti h}}  
  \newcommand{\Ib}{{\uwti I}}

\newcommand{\Ac} {{\mathcal A}}         
\newcommand{\Bc} {{\mathcal B}}         
         \newcommand{\Ck} {{\bm {\mathcal C}}}
\newcommand{\Dc} {{\mathcal D}}         
\newcommand{\Ec} {{\mathcal E}}         
\newcommand{\Fc} {{\mathcal F}}         
\newcommand{\Gc} {{\mathcal G}}         
         
\newcommand{\Ic} {{\mathcal I}}

\newcommand{\Mc} {{\mathcal M}}         \newcommand{\Mk} {{\bm {\mathcal M}}}

\newcommand{\Sc} {{\mathcal S}}         
         
\newcommand{\Uc} {{\mathcal U}}         \newcommand{\Uk} {{\bm {\mathcal U}}}
\newcommand{\Vc} {{\mathcal V}}         \newcommand{\Vk} {{\bm {\mathcal V}}}
         
\newcommand{\Xc} {{\mathcal X}}

\newcommand{\ba}{\begin{array}}
\newcommand{\ea}{\end{array}}

\newtheorem{corollary}{Corollary}
\newtheorem{remark}{Remark}
\newtheorem{theorem}{Theorem}
\newtheorem{lemma}{Lemma}
\newtheorem{proposition}{Proposition}

\newtheorem{assumption}{Assumption}

\begin{document}


%
%
%
%

\lhead{\footnotesize {This article has been accepted for publication in a future issue of this journal, but has not been fully edited. Content may change prior to final publication. IEEE TRANSACTIONS ON MOBILE COMPUTING} }

\title{Multi-User  Scheduling in the 3GPP LTE Cellular Uplink}
\author{
\authorblockN{
Narayan Prasad$^*$, Honghai Zhang$^o$, Hao Zhu$^+$
   and
Sampath Rangarajan$^*$}\thanks{The material in this paper was presented in part at the 10th
International Symposium on Modeling and Optimization in
Mobile, Ad Hoc, and Wireless Networks (WiOpt) 2012.} \thanks{Digital Object Indentifier $10.1109/{\rm TMC}.2012.230$ \hspace{2cm}$1536-1233/12/\$31.00$ \copyright 2012 IEEE}

\IEEEauthorblockA{\small$^*$ NEC Labs America, Princeton; $^o$ Google, Seattle; $^+$ University  of Illinois at Urbana Champaign;\\
 \{prasad,sampath\}@nec-labs.com; honghaiz@google.com; haozhu@illinois.edu
}}
\date{}
\maketitle

\begin{abstract}
In this paper, we consider resource allocation in   the 3GPP Long Term Evolution (LTE)  cellular uplink,  which will be the most widely deployed next generation cellular uplink. The key features of the 3GPP LTE  uplink  (UL) are that it is based on a modified form of the orthogonal frequency division multiplexing based multiple acess (OFDMA) which enables channel dependent frequency selective scheduling, and that it allows for multi-user (MU) scheduling wherein multiple users can be assigned the same time-frequency resource. In addition to the considerable spectral efficiency improvements that are possible by exploiting these two features, the LTE UL allows for transmit antenna selection together with the possibility to employ advanced receivers at the base-station, which promise further gains. However, several practical constraints that seek to  maintain a low signaling overhead, are also imposed. In this paper, we show that the resulting resource allocation problem is   APX-hard and then propose a {\em local ratio test (LRT)} based constant-factor polynomial-time   approximation  algorithm.
We then propose two  enhancements to this algorithm as well as a sequential LRT based MU scheduling algorithm  that offers a constant-factor approximation and is another useful choice in the complexity versus performance tradeoff.  Further, user pre-selection, wherein  a smaller pool of {\em good} users is pre-selected and a sophisticated scheduling algorithm is then employed on the selected pool, is also examined.   We suggest several such user pre-selection algorithms, some of which are shown to offer constant-factor approximations to the pre-selection problem. Detailed evaluations reveal that the proposed algorithms and their enhancements   offer  significant gains.



\end{abstract}


\section{Introduction}
The next generation cellular systems, a.k.a. 4G cellular systems, will operate over wideband multi-path fading channels and have chosen  OFDMA   as their air-interface  \cite{3gpp:rel8}. The motivating factors behind the choice of OFDMA are that it is
an effective means to handle multi-path fading
  and that it allows for enhancing multi-user diversity gains via channel-dependent frequency-domain scheduling.
 The deployment of 4G cellular systems has begun and will accelerate in the coming years. Predominantly the 4G cellular systems will be based on the 3GPP LTE standard \cite{3gpp:rel8} since an overwhelming majority of cellular operators have committed to LTE and specifically all deployments in the forseeable future will adhere to the first version of the LTE standard, referred to as Release 8.  
 Our focus in this paper is on the uplink (UL) in these Release 8 LTE based cellular systems (henceforth referred to simply as LTE UL) and in particular on multi-user (MU) scheduling for the   LTE UL. The  LTE UL employs a modified form of OFDMA, referred to as the DFT-Spread-OFDMA \cite{3gpp:rel8}. In each scheduling interval, the available system bandwidth is partitioned among multiple resource blocks (RBs), where each RB represents the minimum allocation unit and is a pre-defined set of consecutive subcarriers and OFDM symbols. The scheduler  is a frequency domain packet scheduler, which in each scheduling interval assigns these RBs to  the individual users.  Anticipating a rapid growth in data traffic, the LTE UL has enabled MU scheduling   along with transmit antenna selection. Unlike single-user (SU) scheduling, a key feature of  MU scheduling is that an RB can be simultaneously assigned to more that one user in the same scheduling interval. MU scheduling is well supported by  fundamental capacity and degrees of freedom based analysis \cite{YuW:degfree,tse:poly} and indeed, its promised gains need to be harvested in order to cater to the ever increasing traffic demands.
 However, several constraints have also been placed by the LTE standard on such MU  scheduling (and the resulting MU transmissions).  These constraints seek to balance the need to  provide scheduling freedom with the need to ensure a low signaling overhead and respect device limitations. The design of an efficient and implementable MU scheduler for the LTE UL is thus an important problem.

\begin{figure}
\centering
\includegraphics[width=0.9\linewidth]
{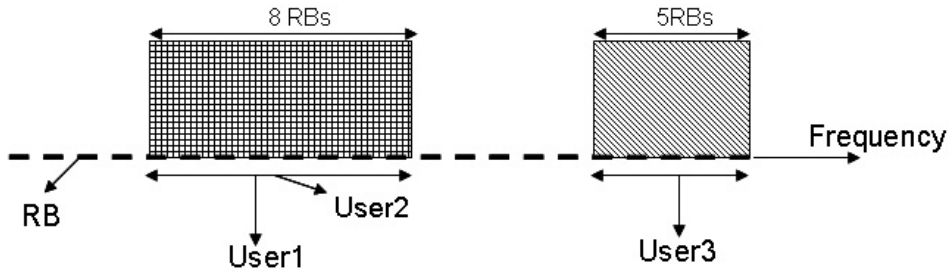} \caption{A Feasible RB Allocation in the LTE UL}
\label{fig:lte}
\end{figure}

 In Fig. \ref{fig:lte} we highlight the key constraints in LTE MU scheduling by depicting a feasible allocation. Notice first that all RBs assigned to a user must form a chunk of contiguous RBs and each user can be assigned at-most one such chunk. This restriction allows us to exploit frequency domain channel variations  via localized assignments (there is complete freedom in choosing the location and size of each such chunk) while respecting strict limits on the per-user transmit peak-to-average-power-ratio (PAPR). Note also that there should be a complete overlap among any two users that share an RB. In other words, if any two users are co-scheduled on an RB then those two users must be co-scheduled on all their assigned RBs.
    This constraint is a consequence  of Zadoff-Chu (ZC) sequences (and their cyclic shifts) being used as pilot sequences   in the LTE UL  \cite{3gpp:rel8} and is needed to ensure reliable channel estimation. 
 The LTE UL further assumes that
  each user can have multiple transmit antennas but is equipped with only one power amplifier due to cost constraints. Accordingly, it allows  a basic precoding in the form of transmit antenna selection where  each scheduled user can be informed about the transmit antenna it should employ in a scheduling interval.
 In addition, to minimize the signaling overhead, each scheduled user can transmit with only one power level (or power spectral density (PSD))  on all its assigned RBs. This PSD is implicitly determined by the number of RBs assigned to that user, i.e., the user divides its total power equally among all its assigned RBs subject possibly to a spectral mask constraint (a.k.a. power pooling). While this constraint significantly decreases the signaling overhead involved in conveying the scheduling decisions to the users, it does not result in any significant performance degradation. This is due to the fact that the multi-user diversity effect ensures that each user is scheduled on the set of RBs on which it has relatively good channels. A constant power allocation over such {\em good} channels results in a negligible loss \cite{YuW:Constant}.
Finally, scheduling in  LTE  UL must respect control channel overhead constraints and interference limit constraints. The former constraints arise because the scheduling decisions are conveyed to the users on the downlink control channel, whose limited capacity in turn places a limit on the set of users that can be scheduled. The latter constraints are employed to mitigate intercell interference. In the sequel it is shown that both these types of constraints can be posed as column-sparse and generic knapsack (linear packing) constraints, respectively.


The goal of this work is to design practical MU resource
allocation algorithms for the LTE   cellular uplink, where the term resource
refers to RBs, modulation and coding schemes (MCS), power levels as well as choice of transmit antennas. In particular, we consider the design of resource allocation algorithms via weighted sum rate utility maximization, which accounts for   finite user queues (buffers)  and practical MCS. In addition, the designed algorithms comply with all the aforementioned practical constraints. 
Our main contributions  are as follows:

 \begin{enumerate}

\item We show that while the {\em complete} overlap constraint along with the at-most one chunk per scheduled user constraint make the resource allocation problem APX-hard, they greatly facilitate the use of local ratio test (LRT) based methods \cite{barnoy:lrt,Yang:ULinfo}.   We then design an LRT based polynomial time deterministic constant-factor approximation algorithm. A remarkable feature of this LRT based algorithm is that it is an end-to-end solution which can accommodate all constraints. 
\item We then propose an enhancement that can significantly reduce the complexity of the LRT based MU scheduling algorithm while offering identical performance, as well as an enhancement that can yield good performance improvements with a very small additional complexity. 

    \item  We propose a sequential LRT based MU scheduling algorithm that offers another useful choice in the complexity versus performance tradeoff. This algorithm also offers constant-factor  approximation (albeit with a poorer constant) and a significantly reduced complexity.

        \item  In a practical system, it is useful to first pre-select a smaller pool of {\em good} users  and then employ a sophisticated scheduling algorithm on the selected pool. Pre-selection can substantially reduce complexity and is also a simple way to enforce a constraint on the number of users that can be scheduled in a scheduling interval. We note that another way to enforce the latter constraint is via a knapsack constraint in the LRT based MU scheduling. We suggest several such user pre-selection algorithms, some of which are shown to offer constant-factor approximations to the pre-selection problem.

            \item The performance of the proposed LRT based MU scheduling algorithm together with its enhancements, the sequential LRT based MU scheduling algorithm and the proposed user pre-selection algorithms are evaluated  for different BS receiver options via elaborate system level simulations that fully conform to the 3GPP evaluation methodology. It is seen that the proposed LRT based MU scheduling algorithm along with an advanced BS receiver can yield over $27\%$ improvement in cell average throughout along with over $10\%$ cell edge throughput improvement compared to SU scheduling. Its sequential counterpart is also attractive in that it yields about $20\%$ improvement in cell average throughput while retaining the  cell edge performance of SU scheduling. Further, it is seen that user pre-selection is indeed an effective approach and the suggested pre-selection approaches can offer significant gains. 

         \end{enumerate}

 \subsection{Related Work}
Resource allocation for the  OFDM/OFDMA networks has been the subject of intense research
 \cite{YuW:dual,YuW:Globe:ofdma,OFDMAJointAllocation,Lee:DLinfo,MPOFDMAICC:2009,lozano:mercury}. A majority of OFDMA resource allocation problems hitherto considered belong to the class of single-user (SU) scheduling problems, which attempt to maximize a system utility by assigning non-overlapping  subcarriers to users, along with transmit power levels for the assigned subcarriers. Even within this class most of the   focus has been on the downlink. These resource allocation problems  have been formulated as {\em continuous optimization problems}, which are in general non-linear and non-convex. As a result
 several approaches
based on the game theory  \cite{OFDMA:Game:Globecom:08,HanZ:nash}, dual decomposition \cite{YuW:dual} or the analysis of optimality conditions \cite{KimandHan:GreedyJointSubcarrierPower:2005} have been developed.
Recent works have focused on the downlink in emerging cellular standards and have proposed approximation algorithms after modeling  the resource allocation problems as constrained integer programs. Prominent examples are \cite{Lee:DLinfo}, \cite{zhangP:2012} which consider the design of downlink SU-MIMO schedulers for LTE cellular systems and derive constant factor approximation algorithms.

Resource allocation for the DFT-Spread-OFDMA uplink has   been relatively  less studied with \cite{multiserver:2009,Yang:ULinfo,Lee-UL-2009,prasad:globe11,madan:2011,OFDMAyang} being the recent examples. In particular, \cite{madan:2011} first considers a relaxed  SU  scheduling problem (without the integer valued RB allocation    and the contiguity constraints) and poses the resource allocation problem as a convex optimization problem. It then proposes a fast interior point based method to solve that problem followed by a modification step to ensure contiguous allocation. A similar approach was adopted earlier in \cite{OFDMAhuang} where the formulated convex optimization problem was solved via a sub-gradient method followed by a modification step to ensure integer valued RB allocation. Furthermore, \cite{OFDMAyang} explicitly enforced the integer valued RB allocation constraint while formulating the resource allocation problem but also assumed that the chunk size for each user is given as an input, and proposed message passing based algorithms. Message passing based algorithms were also applied in \cite{MPOFDMAICC:2009} over an OFDMA uplink in order to minimize the
total transmit power subject to rate guarantees. We note that while the algorithms in \cite{madan:2011,OFDMAhuang,OFDMAyang} may yield effective solutions in different regimes, they do not offer a worst-case performance guarantee and hence cannot be claimed to be approximation algorithms.

On the other hand, \cite{multiserver:2009,Yang:ULinfo,Lee-UL-2009,prasad:globe11} have explicitly modeled both integer valued RB allocation    and the contiguity constraints. Specifically,
\cite{multiserver:2009} shows  that the SU LTE UL  scheduling problem is APX-hard and both \cite{multiserver:2009,Yang:ULinfo} provide  deterministic constant-factor approximation algorithms, whereas \cite{Lee-UL-2009}
provides  a randomized constant-factor approximation algorithm.
\cite{prasad:globe11} extends the algorithms of \cite{multiserver:2009,Yang:ULinfo} to the SU-MIMO LTE-A scheduling.  The algorithm proposed in \cite{Yang:ULinfo} is based on an innovative application of the LRT technique, which was
 developed earlier in \cite{barnoy:lrt}. However, we emphasize that   the algorithms in  \cite{multiserver:2009,Yang:ULinfo,prasad:globe11,Lee-UL-2009}   cannot incorporate MU scheduling,  do not consider user pre-selection  and also  cannot incorporate knapsack constraints. To the best of our knowledge the design of approximation algorithms for MU scheduling in the LTE uplink has not been considered before.

\section{MU Scheduling in the  LTE UL}

Consider a single-cell with $K$ users and one BS which is assumed to have $N_r\geq 1$ receive antennas. Suppose that user $k$ has $N_t\geq 1$ transmit antennas and its power budget is $P_k$.  
We let  $N$ denote the total number of RBs. 

We consider the problem of scheduling users in the frequency domain in a given scheduling interval. 
Let $\alpha_k,  \;1\leq k\leq K$ denote the non-negative weight of the $k^{th}$ user which is an input to the scheduling algorithm and is updated using the output of the scheduling algorithm in every scheduling interval, say according to the proportional fairness rule \cite{Liu-Knightly-2003}. Letting $r_k$ denote the rate assigned to the $k^{th}$ user (in bits per frame of N RBs), we consider the following weighted sum rate utility maximization problem,
\begin{eqnarray}\label{eq:original}
 \max\sum_{1\leq k\leq K}\alpha_{k}r_k,\;\;
\end{eqnarray}
where the maximization is over the assignment of resources to the users \textbf{subject to:}


\begin{itemize}

 \item\textbf{Decodability constraint:} The rates assigned to the scheduled users should be decodable by the base-station receiver. Notice that unlike SU scheduling, MU scheduling allows for multiple users to be assigned the same RB. As a result the rate that can be achieved for user $k$ need not be only a function of the resources assigned to the $k^{th}$ user but can also depend on the those assigned to the other users as well. 

\item \textbf{One transmit antenna and one power level per user:} Each user can transmit using only one power amplifier due to cost constraints. Thus, only a basic precoding in the form of transmit antenna selection is possible. In addition, each scheduled user must perform power pooling, i.e., it is allowed to transmit with only one power level (or power spectral density (PSD))  on all its assigned RBs, where the PSD is implicitly determined by the number of RBs assigned to that user. 

\item \textbf{At most one chunk per-user and at-most $T$ users per RB:} The set of RBs assigned to each scheduled user should form one chunk, where each chunk is a set of contiguous RBs.  Further at-most $T$ users can be co-scheduled on a given RB.  $T$ is expected to be small number typically two.

  \item  \textbf{Complete overlap constraint:} If any two users are assigned a common RB then those two users must be assigned the same set of RBs. 
    Feasible RB allocation and co-scheduling of users in  LTE MU UL  is depicted in Fig \ref{fig:lte}.

 \item\textbf{Finite buffers and finite MCS:} Users in a practical UL will have bursty traffic which necessitates considering finite buffers. In addition, only a finite set of MCS (29 possibilities in the LTE network) can be employed.

\item\textbf{Control channel overhead constraints:} Every user that is given an UL grant (i.e., is scheduled on at least one RB) must be informed about its assigned MCS and the set of RBs on which it must transmit along with possibly the transmit antenna it should employ. This information is sent on the DL control channel of limited capacity which imposes a limit on the set of users that can be scheduled. In particular, the scheduling information of a user is encoded and formatted into one packet (henceforth referred to as a control packet), where the size of the control packet must be selected from a predetermined set of sizes.   A longer (shorter) control packet is used for a cell edge (cell interior) user. In the LTE systems each user is assigned one search region when it enters the cell. In each scheduling interval it then searches for the control packet (containing the scheduling decisions made for it) only in that region of the downlink control channel, as well as a region common to all users.  A more elaborate description is given in the Appendix.

\item\textbf{Per sub-band interference limit constraints:} Inter-cell interference mitigation  is performed 
    by imposing interference limit constraints. In particular, on one or more subbands, the cell of interest must ensure that the total   interference imposed by its scheduled users on a neighboring base-station is below  a specified limit.
 \end{itemize}

We define the set $\Ck$ as the set  containing $N$ length vectors such that any $\cb\in\Ck$ is binary-valued with ($\{0,1\}$) elements and contains a contiguous sequence of ones with the remaining elements being zero. Here  we say an RB $i$ belongs to $\cb$ ($i\in\cb$) if $\cb$ contains a one in its $i^{th}$ position, i.e., $c(i)=1$. Note then that each
  $\cb\in\Ck$ denotes a valid assignment of RBs since it contains one contiguous chunk of RBs.  Also $\cb_1$ and $\cb_2$ are said to intersect if there is some RB that belongs to both $\cb_1$ and $\cb_2$.
For any $\cb\in\Ck$, we  will use ${\rm Tail}(\cb)$ (${\rm Head}(\cb)$) to return the largest (smallest)   index that contains a one in $\cb$. Thus, each $\cb\in\Ck$ has ones in all positions ${\rm Head}(\cb),\cdots,{\rm Tail}(\cb)$ and zeros elsewhere.
Further, we define $\{\Gc_1,\cdots,\Gc_L\}$ to be a partition of $\{1,\cdots,K\}$ with the understanding that all distinct users that belong to a common set (or group) $\Gc_s$, for any $1\leq s\leq L$, are mutually incompatible. In other words at-most one user from each group $\Gc_s$ can be scheduled in a scheduling interval. Notice that by choosing $L=K$ and $\Gc_s=\{s\},\;1\leq s\leq K$ we obtain the case where all users are mutually compatible. Let us define a  family of subsets, $\Uk$, as
 \begin{eqnarray}
  \Uk=\left\{\Uc\subseteq\{1,\cdots,K\}:|\Uc|\leq T\;\&\;|\Uc\cap\Gc_s|\leq 1\;\forall\;1\leq s\leq L\right\}
 \end{eqnarray}
and let $\Mk=\Uk\times\Ck$.

We can now pose the resource allocation problem as
\begin{eqnarray*}\label{eq:RAlte}
 \nonumber \max \sum_{(\Uc,\cb)\in\Mk}p(\Uc,\cb)\Xc(\Uc,\cb),\;\;{\rm s.t.}\\
 \nonumber {\rm For\;\; each \;\;group}\;\;\Gc_s,\;\;\sum_{(\Uc,\cb)\in\Mk\atop \Uc:\Uc\cap\Gc_s\neq\phi} \Xc(\Uc,\cb)\leq 1;\\
 \nonumber {\rm For \;\;each\;\; RB}\;\;i,\;\;\sum_{(\Uc,\cb)\in\Mk\atop \cb:i\in\cb} \Xc(\Uc,\cb)\leq 1;\\
 \nonumber \sum_{(\Uc,\cb)\in\Mk}\beta^q( \Uc,\cb) \Xc(\Uc,\cb)\leq 1,\;1\leq q\leq J; \\
  \sum_{(\Uc,\cb)\in\Mk}\alpha^q( \Uc,\cb) \Xc(\Uc,\cb)\leq 1,\; q\in\Ic,\;\;\;\qquad \qquad\qquad{\rm (P1)}
 \end{eqnarray*}
where $\phi$ denotes the empty set and  $\Xc(\Uc,\cb)$ is an indicator function that returns one if users in $\Uc$ are co-scheduled on the chunk indicated by $\cb$. Note that the first constraint ensures that at-most one user is scheduled from each group and that each scheduled user is assigned at-most one chunk. In addition this constraint also enforces the complete overlap constraint.
The second constraint enforces non-overlap among the assigned chunks. Note that
$p(\Uc,\cb)$ denotes the weighted sum-rate obtained upon co-scheduling the users in $\Uc$   on the chunk indicated by $\cb$. We emphasize that {\em there is complete freedom with respect to the computation of $p(\Uc,\cb)$. Indeed,  it can accommodate finite buffer and practical MCS constraints, account for any particular receiver employed by the BS and  can also incorporate any rule to assign a transmit antenna and a power level to each user in $\Uc$ over the chunk $\cb$.} Clearly, computation of these metrics requires that all channel estimates are available to the BS. In this paper we do not consider   channel estimation related issues (cf. \cite{mehtaN:2007} which considers training in conjunction with antenna selection) and simply assume that reliable estimates are available at the BS to compute all metrics.


The first set of $J$ knapsack constraints in (P1), where $J$ is arbitrary but fixed, are generic knapsack constraints. Without loss of generality, we assume that the weight of the pair $(\Uc,\cb)$ in the $q^{th}$ knapsack, $\beta^q( \Uc,\cb)$,  lies in the interval $[0,1]$. 
Notice that we can simply drop each vacuous constraint, i.e., each constraint $q$ for which  $\sum_{(\Uc,\cb)\in\Mk}\beta^q( \Uc,\cb)\leq 1$.
The second set of knapsack constraints are {\em column-sparse binary knapsack constraints}. In particular, for each pair $ (\Uc,\cb)\in\Mk$ and $q\in\Ic$ we have that $\alpha^q( \Uc,\cb)\in\{0,1\}$. Further, we have that  for each $ (\Uc,\cb)\in\Mk$, $\sum_{q\in\Ic}\alpha^q( \Uc,\cb) \leq \Delta$, where $\Delta$ is arbitrary but fixed and denotes the column-sparsity level. Note that here the cardinality of $\Ic$ can scale polynomially in $KN$ keeping $\Delta$ fixed.

Together these two sets of knapsack constraints can enforce a variety of practical constraints, including the control channel and the interference limit constraints. For instance, defining a generic knapsack constraint as $\beta^1(\Uc,\cb)=\frac{|\Uc|}{\tilde{K}},\;\forall\;(\Uc,\cb)\in\Mk$, for any given input $\tilde{K}$ can enforce that no more that
$\tilde{K}$ can be scheduled in a given interval, which represents a coarse control channel constraint.
In a similar vein, consider any given choice of  a victim adjacent base-station and a sub-band with the constraint that the total interference caused to the victim BS by users scheduled in the cell of interest, over all the RBs in the subband, should be no greater than a specified upper bound.
This constraint can readily modeled using a generic knapsack constraint where the weight of each pair $(\Uc,\cb)\in\Mk$ is simply the ratio of the total interference caused by users in $\Uc$ to the victim BS over RBs that are in $\cb$ as well as the specified subband, and the specified upper bound. The interference is computed using the transmission parameters (such as the power levels, transmit antennas etc) that yield the metric  $p(\Uc,\cb)$. A finer modeling of the LTE control channel constraints is more involved  since it needs to employ the column-sparse knapsack constraints together with the user incompatibility constraints and is deferred to the Appendix. 


Note that for a given $K,N$, an instance of the problem in (P1) consists of a finite set $\Ic$ of indices, a partition $\{\Gc_1,\cdots,\Gc_L\}$, metrics $\{p(\Uc,\cb)\}\;\forall\;(\Uc,\cb)\in\Mk$ and  weights $\{\beta^q( \Uc,\cb)\},\;\forall\;(\Uc,\cb)\in\Mk,1\leq q\leq J$ and $\{\alpha^q( \Uc,\cb)\},\;\forall\;(\Uc,\cb)\in\Mk, q\in\Ic$.
Then, in order to  handle the generic knapsack constraints, we leverage the idea developed in \cite{barnoy:lrt} and first partition the set $\Mk$ into two parts as $\Mk=\Mk^{\rm narrow}\cup \Mk^{\rm wide}$, where we define $\Mk^{\rm narrow}=\{(\Uc,\cb)\in\Mk:\beta^q( \Uc,\cb)\leq 1/2,\;\forall\;1\leq q\leq J\}$ so that $\Mk^{\rm wide}=\Mk\setminus\Mk^{\rm narrow}$.
 We then define $J$ sets, $\Vk^{(1)},\cdots,\Vk^{(J)}$ that cover $\Mk^{\rm wide}$ (note that any two of these sets can mutually overlap) as
   $(\Uc,\cb)\in\Vk^{(q)}\;{\rm iff}\;\beta^q( \Uc,\cb)>1/2$ for $q=1,\cdots,J$. Recall that $T,J$ are fixed and  note that the cardinality of $\Mk$, $|\Mk|$, is $O(K^TN^2)$ and that $\Mk^{\rm narrow}$ and $\{\Vk^{(q)}\}$ can be determined in polynomial time.
Next,   we
propose Algorithm I whose complexity   is essentially determined by that of its module Algorithm IIa and scales polynomially in $KN$ (recall that $T$ is a constant).   
A detailed discussion on the complexity along with steps to reduce it are deferred to the next section.
We offer the following theorem. 
\begin{theorem}\label{thm:lteS}
The problem in (P1) is APX-hard, i.e., there is an $\epsilon>0$ such that it is NP hard to obtain a $1-\epsilon$ approximation algorithm for (P1). Let $\hat{W}^{\rm opt}$ denote the optimal weighted sum rate obtained upon solving (P1) and let $\hat{W}$ denote the   weighted sum rate obtained upon using Algorithm I. Then, we have that
\begin{eqnarray}\label{eq:pRAlte1}
  \hat{W}\geq \left\{\begin{array}{c}\frac{\hat{W}^{\rm opt}}{1+T+\Delta+2J},\;\;{\rm If}\;\;\Mk^{\rm wide}=\phi \\
  \frac{\hat{W}^{\rm opt}}{1+T+\Delta+3J},\;\;{\rm Otherwise}\end{array}\right.
 \end{eqnarray}
\end{theorem}

\proof Let us specialize (P1) to instances where all the knapsack constraints are vacuous, where $L=K$ and $\Gc_s=\{s\},\;1\leq s\leq K$  and where $p(\Uc,\cb)=0$ whenever $|\Uc|\geq 2$ for all $(\Uc,\cb)\in\Mk$. Then (P1) reduces to the SU scheduling problem considered in \cite{Yang:ULinfo,multiserver:2009} which was shown there to be APX-hard. Consequently, we can assert that (P1) is APX-hard.

Next, consider first Algorithm IIa which outputs a feasible allocation over $\Mk^{\rm narrow}$ yielding a weighted sum rate $\hat{W}^{\rm narrow}$. Let $\hat{W}^{\rm opt,narrow}$ denote the optimal weighted sum rate obtained by solving (P1) albeit where  all pairs $(\Uc,\cb)$ are restricted to lie in $\Mk^{\rm narrow}$.
 In Proposition I given in the Appendix, we prove that
 \begin{eqnarray}\label{eq:pRAlte2}
  \hat{W}^{\rm narrow}\geq
  \frac{\hat{W}^{\rm opt,narrow}}{1+T+\Delta+2J}.
 \end{eqnarray}
Our proof  (given in the Appendix) invokes  notation and results developed for LRT based SU scheduling in  \cite{Yang:ULinfo} as much as possible, and highlights mainly the key differences. These differences are novel and crucial since they allow us to co-schedule multiple users on a chunk while respecting incompatibility constraints and to satisfy multiple knapsack constraints.

Next, let us  consider the remaining part which arises when $\Mk^{\rm wide}\neq\phi$.
Consider first Algorithm IIb which outputs a feasible allocation over $\Mk^{\rm wide}$ yielding a weighted sum rate $\hat{W}^{\rm wide}$. Let $\hat{W}^{\rm opt,wide}$ denote the optimal weighted sum rate obtained by solving (P1) albeit where  all pairs $(\Uc,\cb)$ are restricted to lie in $\Mk^{\rm wide}$.
 We will prove that
 \begin{eqnarray}\label{eq:pRAlte8}
  \hat{W}^{\rm wide}\geq
  \frac{\hat{W}^{\rm opt,wide}}{J}.
 \end{eqnarray}
Let $\Vc ^{\rm opt,wide}$ be an optimal allocation of pairs from $\Mk^{\rm wide}$ that results in a weighted sum rate $\hat{W}^{\rm opt,wide}$. Clearly, in order to meet the knapsack constraints,
 $\Vc ^{\rm opt,wide}$ can include at-most one pair from each $\Vk^{(q)},\;1\leq q\leq J$ so that there can be at-most $J$ pairs in  $\Vc ^{\rm opt,wide}$.
 Thus, by selecting the pair yielding the maximum weighted sum-rate we can achieve at-least $\hat{W}^{\rm opt,wide}/J$. The greedy algorithm first selects  the pair yielding the maximum weighted sum rate among all pairs in $\Mk^{\rm wide}$ and then attempts to add pairs to monotonically improve the objective. Thus, we can conclude that (\ref{eq:pRAlte8}) must be true. \endproof

 Notice that we select $\hat{W}=\max\{\hat{W}^{\rm narrow},\hat{W}^{\rm wide}\}$ so that
 \begin{eqnarray}\label{eq:pRAlte8b}
  \hat{W}\geq  \max\left\{\frac{\hat{W}^{\rm opt, narrow}}{1+T+\Delta+2J},\frac{\hat{W}^{\rm opt, wide}}{J}\right\}.
 \end{eqnarray}
 It is readily seen that
 \begin{eqnarray}\label{eq:pRAlte8c}
   \hat{W}^{\rm opt} \leq \hat{W}^{\rm opt, narrow} + \hat{W}^{\rm opt, wide}.
 \end{eqnarray}
 (\ref{eq:pRAlte8b}) and (\ref{eq:pRAlte8c}) together prove the theorem. \endproof
For clarity, all the important symbol definitions are captured in Table \ref{tab:defnsym}.

 An interesting observation that follows from the proof of Theorem \ref{thm:lteS} is that any optimal allocation over $\Mk^{\rm wide}$ can include at-most one pair from each $\Vk^{(q)},\;1\leq q\leq J$. Then since the number of pairs in each   $\Vk^{(q)},\;1\leq q\leq J$ is $O(K^TN^2)$, we can determine an optimal allocation yielding $\hat{W}^{\rm opt,wide}$ via exhaustive enumeration with a high albeit polynomial complexity (recall that $T$ and $J$ are assumed to be fixed). Thus, by using exhaustive enumeration instead of Algorithm IIb, we can claim the following result.
\begin{corollary}
Let $\hat{W}^{\rm opt}$ denote the optimal weighted sum rate obtained upon solving (P1) and let $\hat{W}$ denote the   weighted sum rate obtained upon using Algorithm I albeit with exhaustive enumeration over $\Mk^{\rm wide}$. Then, we have that
\begin{eqnarray}
  \hat{W}\geq \left\{\begin{array}{c} \frac{\hat{W}^{\rm opt}}{1+T+\Delta+2J},\;\;{\rm If}\;\;\Mk^{\rm wide}=\phi\\
  \frac{\hat{W}^{\rm opt}}{2+T+\Delta+2J},\;\;{\rm Otherwise}\end{array}\right.
 \end{eqnarray}
\end{corollary}
\algsetup{indent=1em}
\begin{table}
\caption{{\bf Algorithm I: Algorithm  for LTE UL MU-MIMO}}\label{algo:lrtlteM}
\begin{algorithmic}[1]
\STATE Input $p(\Uc,\cb),\;\;\forall\;(\Uc,\cb)\in\Mk$ and $\Mk^{\rm narrow}, \Mk^{\rm wide}$
\STATE Determine a feasible allocation over $\Mk^{\rm narrow}$ using Algorithm IIa and let $\hat{W}^{\rm narrow}$ denote the corresponding weighted sum rate.
\STATE Determine  a feasible allocation over $\Mk^{\rm wide}$ using Algorithm IIb and let $\hat{W}^{\rm wide}$ denote the corresponding weighted sum rate.
\STATE Select and output the allocation resulting in  $\hat{W}=\max\{\hat{W}^{\rm narrow},\hat{W}^{\rm wide}\}$.
\end{algorithmic}
\end{table}

\algsetup{indent=1em}
\begin{table}
\caption{{\bf Algorithm IIa: LRT based module $\Mk^{\rm narrow}$ }}\label{algo:lrtlte}
\begin{algorithmic}[1]
\STATE Initialize $p'(\Uc,\cb)\leftarrow p(\Uc,\cb),\;\forall\;\;(\Uc,\cb)\in\Mk^{\rm narrow}$, stack $\Sc=\phi$
\STATE \textbf{For} $j=1,\cdots,N$
\STATE Determine $(\Uc^*,\cb^*)=\arg\max_{(\Uc,\cb)\in\Mk^{\rm narrow}\atop {\rm Tail}(\cb)=j}p'(\Uc,\cb)$
\STATE  \textbf{If} $p'(\Uc^*,\cb^*)>0$ \textbf{Then}
\STATE Set $\hat{p}=p'(\Uc^*,\cb^*)$ and ${\rm Push}\; (\Uc^*,\cb^*)\; {\rm into}\; \Sc$.
\STATE \textbf{For} each  $(\Uc,\cb)\in\Mk^{\rm narrow}$ such that $p'(\Uc,\cb)>0$
\STATE  \textbf{If} $ \exists\;\Gc_s:\Uc\cap\Gc_s\neq\phi\;\&\;\Uc^*_j\cap\Gc_s\neq \phi$ or $ \cb^*\cap\cb\neq \phi$ \textbf{Then}
\STATE Update $p'(\Uc,\cb)\leftarrow p'(\Uc,\cb)-\hat{p}$
\STATE  \textbf{Else If}$\;\exists\;q\in\Ic:\alpha^q( \Uc,\cb)=\alpha^q(\Uc_j^*,\cb_j^*)=1$ \textbf{Then}
\STATE Update $p'(\Uc,\cb)\leftarrow p'(\Uc,\cb)-\hat{p}$
\STATE \textbf{Else}
\STATE Update $p'(\Uc,\cb)\leftarrow p'(\Uc,\cb)-2\hat{p}\max_{1\leq q\leq J}\beta^q( \Uc,\cb)$.
    \STATE \textbf{End If}
\STATE  \textbf{End For}
\STATE \textbf{End If}
\STATE  \textbf{End For}
\STATE Set stack $\Sc'=\phi$
\STATE \textbf{While} $\Sc\neq\phi$
\STATE  Obtain $(\Uc,\cb)={\rm Pop}\;\Sc$
 \STATE  \textbf{If}    $(\Uc,\cb)\cup\Sc'$ is valid \textbf{Then}\;\;\;\; \%\%  {\small $(\Uc,\cb)\cup\Sc'$ {\rm {\em is deemed valid if no user in $\Uc$ is incompatible with any user present in $\Sc'$  and no  chunk in $\Sc'$ has an overlap with $\cb$ and all knapsack constraints are satisfied by $(\Uc,\cb)\cup\Sc'$}}}.
\STATE Update $\Sc'\leftarrow (\Uc,\cb)\cup\Sc'$
 \STATE \textbf{End While}
\STATE Output $\Sc'$ and $\hat{W}^{\rm narrow}=\sum_{(\Uc,\cb)\in\Sc'}p(\Uc,\cb)$.
\end{algorithmic}
\end{table}

\algsetup{indent=1em}
\begin{table}
\caption{{\bf Algorithm IIb: Greedy module over $\Mk^{\rm wide}$ }}\label{algo:lrtlte}
\begin{algorithmic}[1]
\STATE Input $p(\Uc,\cb),\;\;\forall\;(\Uc,\cb)\in\Mk^{\rm wide}$ and $\{\Vk^{(q)}\}_{q=1}^J$.
\STATE Set $\Sc=\phi$ and $\Mk'=\Mk^{\rm wide}$.
\STATE \textbf{Repeat}
\STATE Determine $(\Uc^*,\cb^*)=\arg\max_{(\Uc,\cb)\in\Mk'\atop \Sc\cup(\Uc,\cb)\;{\rm is\;valid}} p(\Uc,\cb)$.
\STATE  Update $\Sc\leftarrow\Sc\cup(\Uc^*,\cb^*)$ and $\Mk'=\Mk'\setminus\{\Vk^{(q)}:(\Uc^*,\cb^*)\in\Vk^{(q)}\}$
\STATE  \textbf{Until} $(\Uc^*,\cb^*)=\phi$ or $\Mk'=\phi$.
\STATE Output $\Sc$ and $\hat{W}^{\rm wide}=\sum_{(\Uc,\cb)\in\Sc}p(\Uc,\cb)$.
\end{algorithmic}
\end{table}

\begin{remark}
Some intuition on the process in the heart of Algorithm I (which is Algorithm IIa) is on order. Note that Algorithm IIa has two stages. The first one (comprising of steps 1 through 16) begins by initializing an empty stack $\Sc$ and defining the current gain of each pair to be equal to its metric. Then, promising pairs are successively added to the top of the stack $\Sc$. Each time a pair is pushed into the stack, the current gain  of each pair  that can potentially be added and which conflicts  with the  pair just added (in terms of sharing a common RB or each having a user that belongs to an identical group or each having a unit weight in a common sparse knapsack constraint in $\Ic$), is decremented by the current gain of the added pair. The idea behind this operation is that eventually only one pair among these conflicting pairs can be selected, so by decrementing the gains we ensure that a conflicting pair can be added in a later step only if it has a larger gain. Similarly, the gain of a non conflicting pair is also decremented by its maximal weight times twice the current gain of the added pair, in order account for the non-sparse knapsack constraints. At the end of the first stage the stack $\Sc$ contains a set of promising pairs but the entire set need not be feasible for (P1). In the second stage another stack $\Sc'$ is formed by successively picking pairs from the top of stack $\Sc$ and adding them to $\Sc'$ if feasibility is satisfied. Note that the top down approach of picking pairs from $\Sc$ is intuitively better since pairs at the top will have larger metrics than pairs below with whom they conflict.
\end{remark}

\begin{table}
\begin{small}
\begin{center}
\hspace{-1cm}\caption[]{Symbol Definitions}\label{tab:defnsym}
\begin{tabular}{|c|p{5cm}|c|p{5cm}|}\hline
   $K$ & Number of users &  $N$ & Number of RBs \\ \hline  $L$ & Number of user groups &
   $N_r$ & Number of receive antennas at BS \\ \hline   $N_t$ & Number of transmit antennas at each user & $T$ & Maximum number of co-scheduled users  \\ \hline
 $\alpha_k$ & Weight of  user $k$   & $r_k$ & rate (bits/frame) of user $k$ \\ \hline   $P_k$ & Power budget of user $k$ &
     $\cb$ & $N$-length vector representing a chunk of RBs \\ \hline  ${\rm Head}(\cb)$ & First RB in chunk $\cb$  & ${\rm Tail}(\cb)$ & Last RB in chunk $\cb$ \\ \hline
     $\Ck$ & Set of all valid chunks & $\Gc_k$ & $k^{th}$ group of mutually incompatible users\\ \hline
     $\Uc$ & user subset containing at most $T$ compatible users & $\Uk$ & Family of all valid user subsets\\  \hline $\Mk=\Uk\times\Ck$ & Family of all feasible pairs $\{(\Uc,\cb)\}$ & $p(\Uc,\cb)$ & weighted sum rate obtained upon scheduling pair $(\Uc,\cb)$\\  \hline
     $\beta^q(\Uc,\cb)$ & weight of  $(\Uc,\cb)$ in  $q^{th}$ generic knapsack constraint &
     $\alpha^q(\Uc,\cb)$ & weight of $(\Uc,\cb)$ in  $q^{th}$ sparse knapsack constraint\\  \hline
     $\Mk^{\rm narrow}$ & All feasible pairs $\{(\Uc,\cb):\beta^q(\Uc,\cb)\leq 1/2\;\forall\;q\}$ & $\Mk^{\rm wide}$ & $=\Mk\setminus\Mk^{\rm narrow}$ \\\hline
          $J$ & Number of generic knapsack constraints & $\Ic$  & Set of indices of sparse knapsack constraints \\  \hline $\Xc(\Uc,\cb)$ & Indicator function for scheduling pair $(\Uc,\cb)$ & $\Gamma^{(j)}(\Uc,\cb)$ & Offset for pair $(\Uc,\cb)$ in the $j^{th}$ iteration \\  \hline
     $p^{\rm mmse}(\Uc,j)$ & weighted sum rate obtained upon scheduling  user set $\Uc$ on RB $j$ with MMSE receiver & $p^{\rm sic}(\Uc,j)$ & weighted sum rate obtained upon scheduling user set $\Uc$ on RB $j$ with SIC receiver\\ \hline
     \end{tabular}
\end{center}
\end{small}
\end{table}

For notational simplicity, henceforth unless otherwise mentioned, we assume that all users are mutually compatible, i.e., $L=K$ with $\Gc_s=\{s\},\;1\leq s\leq K$.
\section{Complexity Reduction}\label{sec:compred}
In this section we present key techniques to significantly reduce the complexity of our proposed local ratio test based multi-user scheduling algorithm. As noted before the complexity of Algorithm I is dominated by that of its component Algorithm IIa. Accordingly, we focus our attention on Algorithm IIa and without loss of generality we assume that $\Mk=\Mk^{\rm narrow}$. We first note that for a given set of  metrics $\{p(\Uc,\cb):(\Uc,\cb)\in\Mk\}$, the complexity (in terms of number of operations) of Algorithm IIa scales as $O(K^TN^3)$, with the underlying operations being simple additions of real valued numbers. However, in practise the $O(K^TN^2)$ many metrics have to be first computed. Notice that the metric of any pair $(\Uc,\cb)$ is in general not separable over the constituent RBs in $\cb$. \footnote{This is  due to the fact that the metric must account for the DFT spreading   which each user must employ over the LTE UL.} Each such metric requires the computation of $|\Uc|({\rm Tail}(\cb)- {\rm Head}(\cb)+1)$ signal-to-noise-ratio (SINR) terms (which involve multiplications of complex numbers and possibly  matrix inversions) as well as evaluating transcendental functions (such as $\ln(.)$). Moreover, the power pooling greatly limits re-using SINR terms even across different metrics involving the same user group $\Uc$. Consequently,  the total metric computation complexity can itself scale as $O(K^TN^3)$ but where the underlying operations are much more complex. As a result,  the metric computation can often be the main bottleneck and indeed must be accounted for.

Before proceeding, we make the following assumption that is satisfied by all physically meaningful metrics.
\begin{assumption}\label{assump1}
{\em Sub-additivity}: We assume that for any $(\Uc,\cb)\in\Mk$
\begin{eqnarray}
  p(\Uc,\cb)\leq p(\Uc_1,\cb) + p(\Uc_2,\cb),\;\;\forall\;\; \Uc_1,\Uc_2:\Uc=\Uc_1\cup\Uc_2.
 \end{eqnarray}
\end{assumption}
The following features can then be exploited for a significant reduction in complexity.
\begin{itemize}
\item {\em On demand metric computation:} Notice in Algorithm IIa that the metric for any $(\Uc,\cb)\in\Mk$, where ${\rm Tail}(\cb)=j$ for some $j=1,\cdots,N$, needs to be computed only at the $j^{th}$ iteration at which point we need to determine
   \begin{eqnarray}\label{eq:uppp}
  p'(\Uc,\cb)= p(\Uc,\cb) - \Gamma^{(j)}(\Uc,\cb),
 \end{eqnarray}
where the offset factor $\Gamma^{(j)}(\Uc,\cb)$ is given by
 \begin{eqnarray}
  \nonumber \Gamma^{(j)}(\Uc,\cb)= \sum_{(\Uc^*_m,\cb^*_m)\in\Sc} \left(\tilde{p}(\Uc^*_m,\cb^*_m)\Ec((\Uc,\cb),(\Uc^*_m,\cb^*_m))   + 2\tilde{p}(\Uc^*_m,\cb^*_m)\max_{1\leq q\leq J}\{\beta^q(\Uc,\cb)\}\Ec^c((\Uc,\cb),(\Uc^*_m,\cb^*_m))\right)
 \end{eqnarray}
and where $\tilde{p}(\Uc^*_m,\cb^*_m)$ is equal to the $p'(\Uc^*_m,\cb^*_m)$ computed for the pair selected at the $m^{th}$ iteration with $m\leq j-1$ and  $\Ec((\Uc,\cb),(\Uc^*_m,\cb^*_m))$ denotes an indicator (with $\Ec^c((\Uc,\cb),(\Uc^*_m,\cb^*_m))=1 -\Ec((\Uc,\cb),(\Uc^*_m,\cb^*_m))$)  which is true when $ \Uc^*_m\cap\Uc\neq \phi\;{\rm or}\; \cb\cap\cb^*_m\neq \phi\;{\rm or}\;\exists\;q\in\Ic:\alpha^q(\Uc^*_m,\cb^*_m)=\alpha^q(\Uc,\cb)=1 $.
Further note that $p'(\Uc,\cb)$ in (\ref{eq:uppp}) is required only if it is strictly positive.
Then,
an important observation is that if at the $j^{th}$ iteration, we have already computed
$p(\Uc_1,\cb)$ and $p(\Uc_2,\cb)$ for some $\Uc_1,\Uc_2:\Uc=\Uc_1\cup\Uc_2$, then invoking the sub-additivity property we have that
 \begin{eqnarray}\label{eq: subaddmetu}
  p'(\Uc,\cb) \leq p(\Uc_1,\cb) + p(\Uc_2,\cb) - \Gamma^{(j)}(\Uc,\cb),
 \end{eqnarray}
so that if the RHS in (\ref{eq: subaddmetu}) is not strictly positive or if it is less than the greatest value of $p'(\Uc',\cb')$ computed in the current iteration for some other pair $(\Uc',\cb'):{\rm Tail}(\cb')=j$, then {\em we do not need to compute $p'(\Uc,\cb)$ and hence the metric $p(\Uc,\cb)$}.

\item {\em Selective update} Note that in the $j^{th}$ iteration, once the best pair $(\Uc^*_j,\cb^*_j)$ is selected and it is determined that $p'(\Uc^*_j,\cb^*_j)>0$, we need to update the metrics for pairs   $(\Uc',\cb'):{\rm Tail}(\cb')\geq j+1$, since only such pairs will be considered in future iterations. Thus, the offset factors $\{\Gamma^{(j)}(\Uc',\cb')\}$ need to be updated only for such pairs, via
    \begin{eqnarray*}
   \hspace{-1cm}\Gamma^{(j+1)}(\Uc',\cb')=\Gamma^{(j)}(\Uc',\cb') +  p'(\Uc^*_j,\cb^*_j)\Ec((\Uc',\cb'),(\Uc^*_j,\cb^*_j))   + 2p'(\Uc^*_j,\cb^*_j)\max_{1\leq q\leq J}\{\beta^q(\Uc',\cb')\}\Ec^c((\Uc',\cb'),(\Uc^*_j,\cb^*_j)).\;\;\;\;\;
 \end{eqnarray*}
    Further, if by exploiting sub-additivity we can deduce that $p'(\Uc',\cb')\leq 0$ for any such pair, then we can drop such a pair along with its offset factor from future consideration.

 \end{itemize}

%
%

\section{Improving Performance via a second phase}\label{sec:SP}
A potential drawback of the LRT based algorithm is that some RBs may remain un-utilized, i.e., they may not be assigned to any user. Notice that when the final stack $\Sc'$ is built in the while-loop of Algorithm IIa, an allocation or pair from the top of stack $\Sc$ is added to stack $\Sc'$ only if it does not violate feasibility when considered together with those already  in stack $\Sc'$. Often multiple pairs from $\Sc$ are dropped due to such feasibility violations,  resulting in spectral holes formed by unassigned RBs. To mitigate this problem, we perform a second phase.
The second phase consists of running Algorithm IIa again albeit with modified metrics $\{\breve{p}(\Uc,\cb):(\Uc,\cb)\in \Mk^{\rm narrow}\}$ which are obtained via the following steps.
 \begin{enumerate}
 \item Initialize $\breve{p}(\Uc,\cb)=p(\Uc,\cb),\;\forall\;(\Uc,\cb)\in \Mk^{\rm narrow}$. Let $\Sc'$ be obtained as the output of Algorithm IIa when it is implemented first.

\item For each $(\Uc,\cb)\in\Sc'$, we ensure that any user in $\Uc$ is not scheduled by phase two in any other user set save $\Uc$, by setting
\begin{eqnarray}
   \breve{p}(\Uc',\cb') = 0 \;{\rm if}\; \Uc'\neq\Uc\;\&\;\Uc'\cap\Uc\neq\phi,\;\forall\;(\Uc',\cb')\in \Mk^{\rm narrow}.
 \end{eqnarray}
 \item For each $(\Uc,\cb)\in\Sc'$, we ensure that no other user set save $\Uc$ is assigned any RB in $\cb$, by setting
\begin{eqnarray}
   \breve{p}(\Uc',\cb') = 0 \;{\rm if}\; \Uc'\neq\Uc\;\&\;\cb'\cap\cb\neq\phi,\;\forall\;(\Uc',\cb')\in \Mk^{\rm narrow}.
 \end{eqnarray}

 \item For each $(\Uc,\cb)\in\Sc'$, we ensure that the allocation $(\Uc,\cb)$ is either unchanged by phase two or is expanded,  by setting
\begin{eqnarray*}
    \breve{p}(\Uc,\cb') &=& \left\{\begin{array}{c} p(\Uc,\cb'),\;\; {\rm If}\;{\rm Tail(\cb')}\geq {\rm Tail(\cb)} \;\&\; {\rm Head(\cb')}\leq {\rm Head(\cb)}\\
  0,\;\;{\rm Otherwise}\end{array}\right.\\
 \end{eqnarray*}
\end{enumerate}
A consequence of using the modified metrics is that the second phase has a significantly less complexity since a large fraction of the allocations are disallowed (since many of the modified metrics are zero). While the second phase does not offer any improvement in the approximation factor, simulation results presented in the sequel reveal that it offers a good performance improvement with very low complexity addition.

\section{Simulation Results: Single cell Setup}

In this section we evaluate key features of our proposed algorithm over an idealized single-cell setup.
 In particular, we simulate an uplink  wherein the BS is equipped with four receive antennas.  We model the fading channel between each user and the BS as a six-path equal gain i.i.d. Rayleigh fading channel and assume an infinitely backlogged traffic model. For simplicity, we assume that there are no knapsack constraints and that at-most two users can be co-scheduled on an RB (i.e., $J=0,\Delta=0$ and $T=2$). Further, each user can employ ideal Gaussian codes and upon being scheduled, divides its maximum transmit power equally among its assigned RBs. Notice that since $\Mk=\Mk^{\rm narrow}$ we can directly use Algorithm IIa.

\begin{figure}[ht]
	\begin{minipage}[b]{0.45\linewidth}
	\centering
	\includegraphics[width=\textwidth]{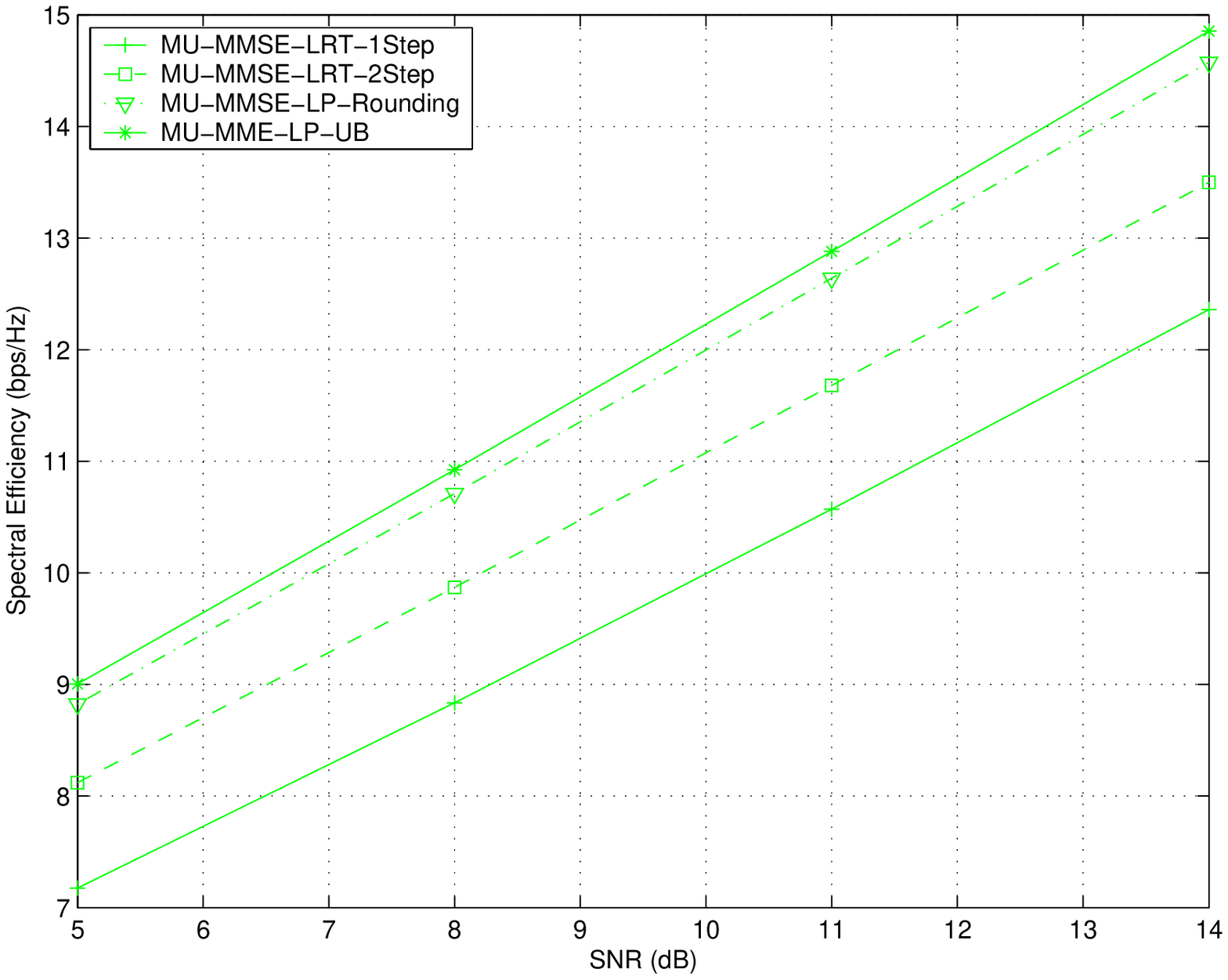}
	\caption{Average spectral efficiency versus SNR (dB):  MU Scheduling with MMSE receiver.}
	\label{fig_plotscmmse}
	\end{minipage}
	\hspace{0.5cm}
	\begin{minipage}[b]{0.45\linewidth}
	\centering
	\includegraphics[width=\textwidth]{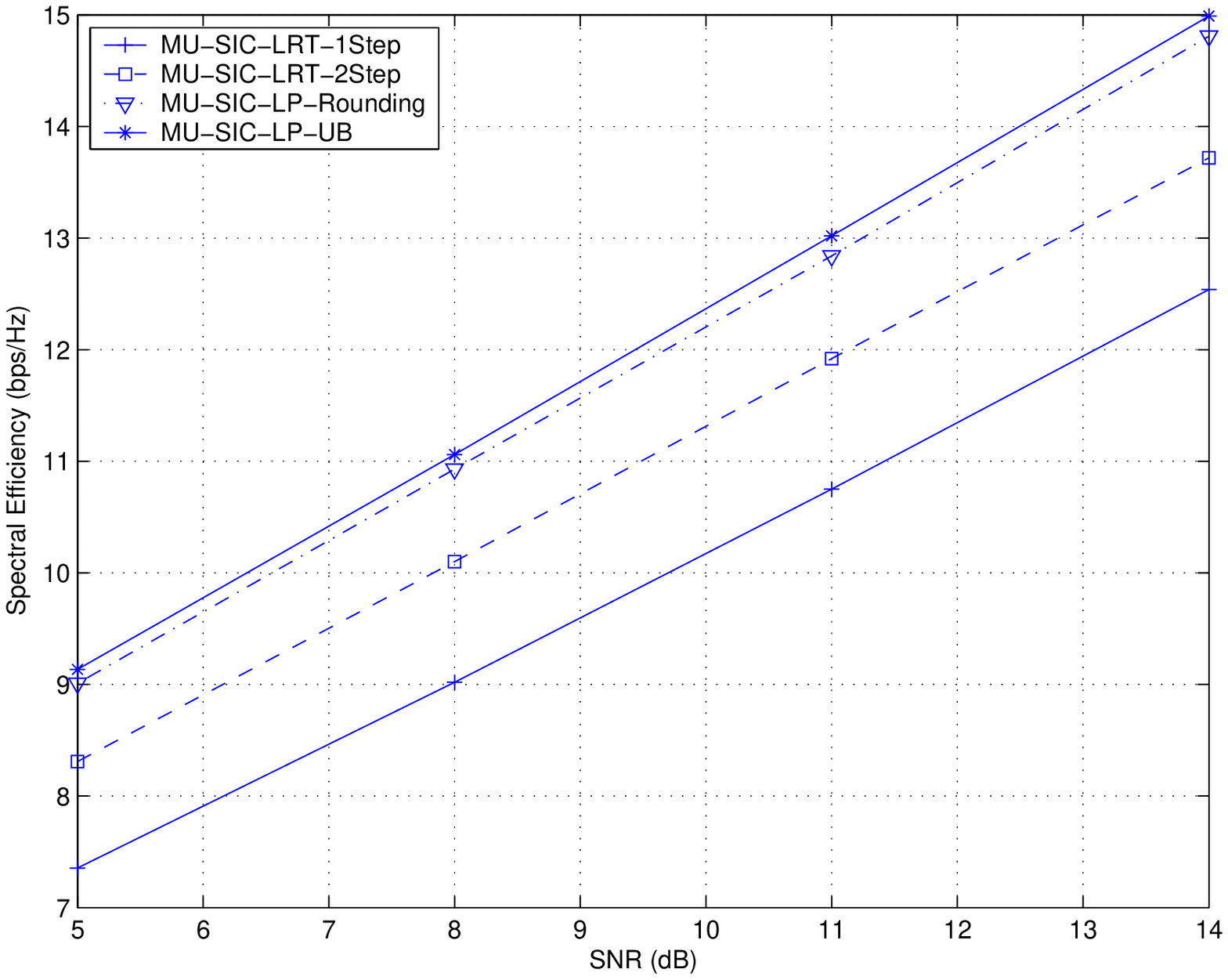}
	\caption{Average spectral efficiency versus SNR (dB):  MU Scheduling with SIC receiver.}
	\label{fig_plotscsic}
	\end{minipage}
	\end{figure}

\begin{figure}[ht]
	\begin{minipage}[b]{0.45\linewidth}
	\centering
	\includegraphics[width=\textwidth]{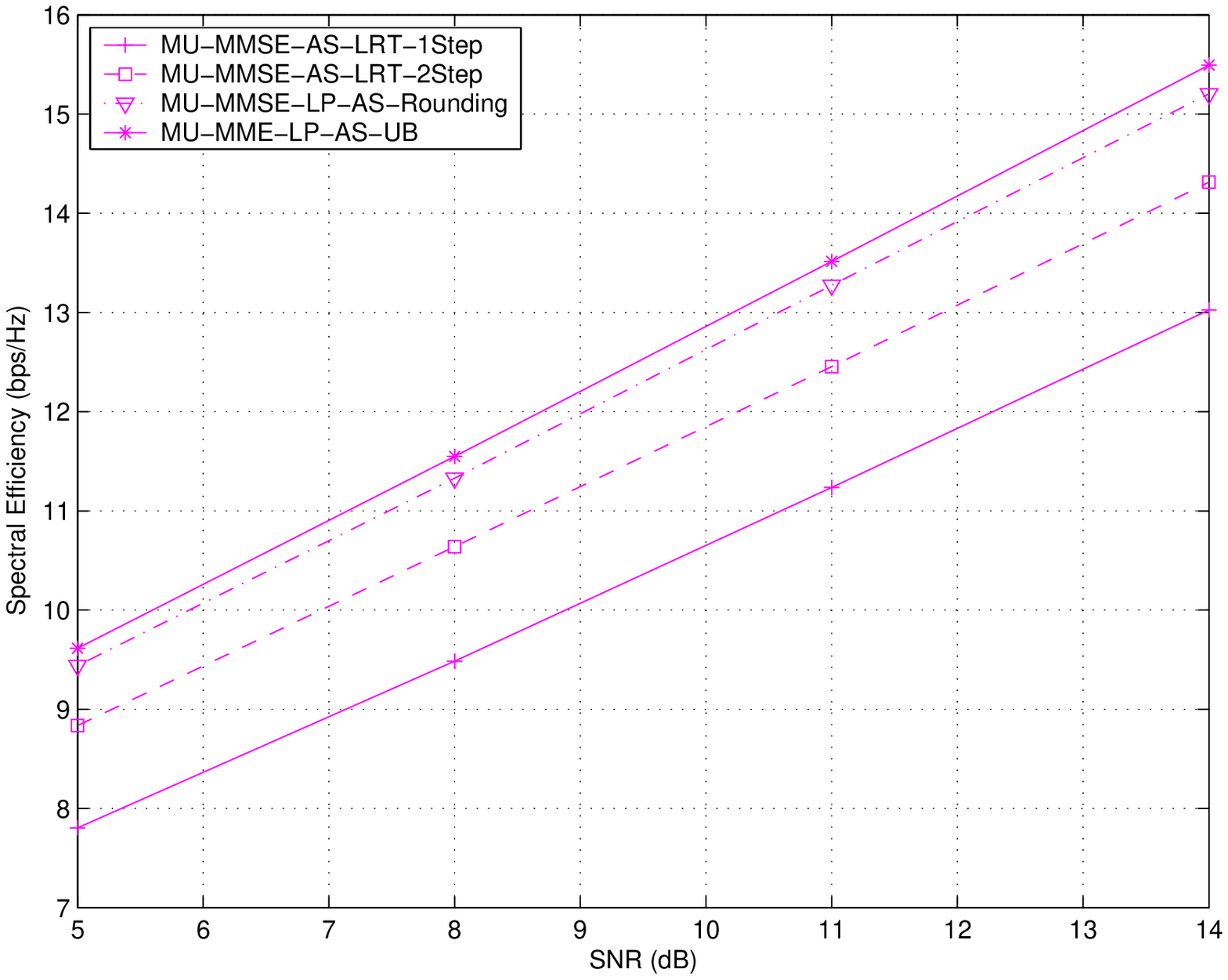}
	\caption{Average spectral efficiency versus SNR (dB):  MU Scheduling with MMSE and Antenna Selection.}
	\label{fig_plotscmmseas}
	\end{minipage}
	\hspace{0.5cm}
	\begin{minipage}[b]{0.45\linewidth}
	\centering
	\includegraphics[width=\textwidth]{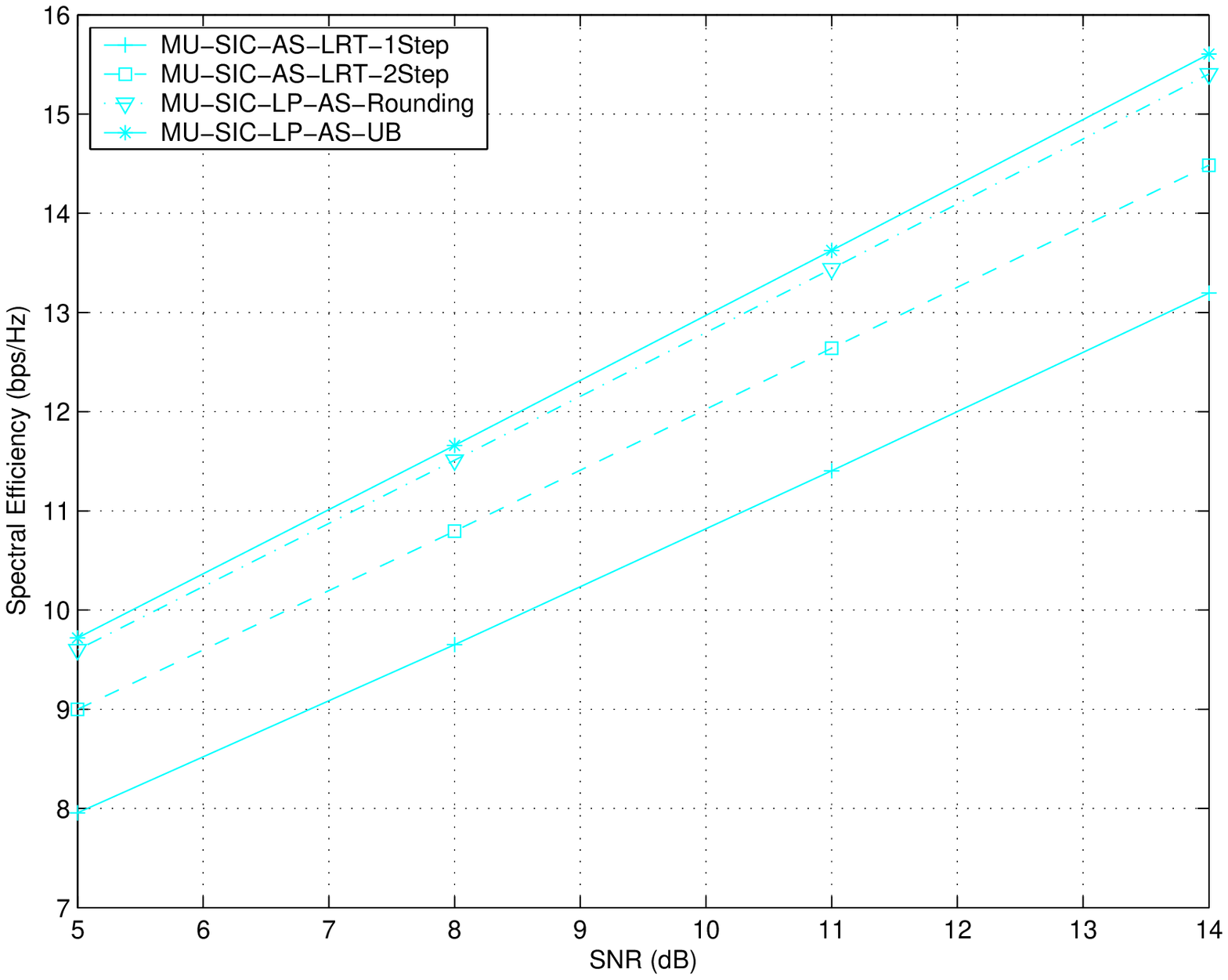}
	\caption{Average spectral efficiency versus SNR (dB): MU Scheduling with SIC and Antenna Selection.}
	\label{fig_plotscsicas}
	\end{minipage}
	\end{figure}

\begin{figure}[!t]
\centering
\includegraphics[width=3.5in]{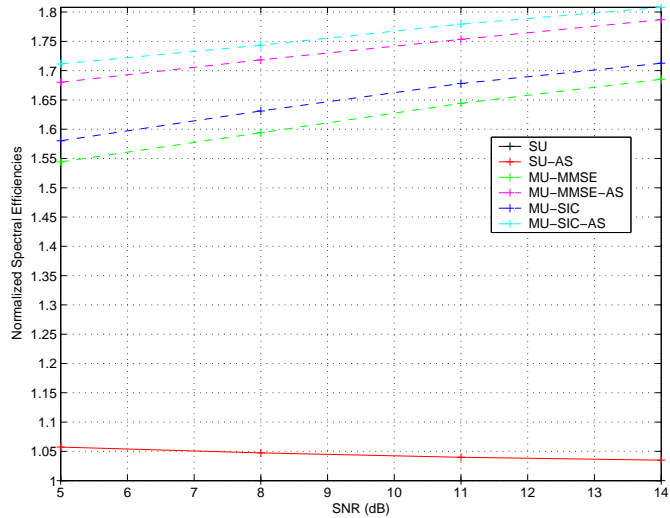}
\caption{Normalized spectral efficiency versus SNR (dB)}
\label{fig_plotnse}
\end{figure}

\begin{figure}[ht]
	\begin{minipage}[b]{0.45\linewidth}
	\centering
	\includegraphics[width=\textwidth]{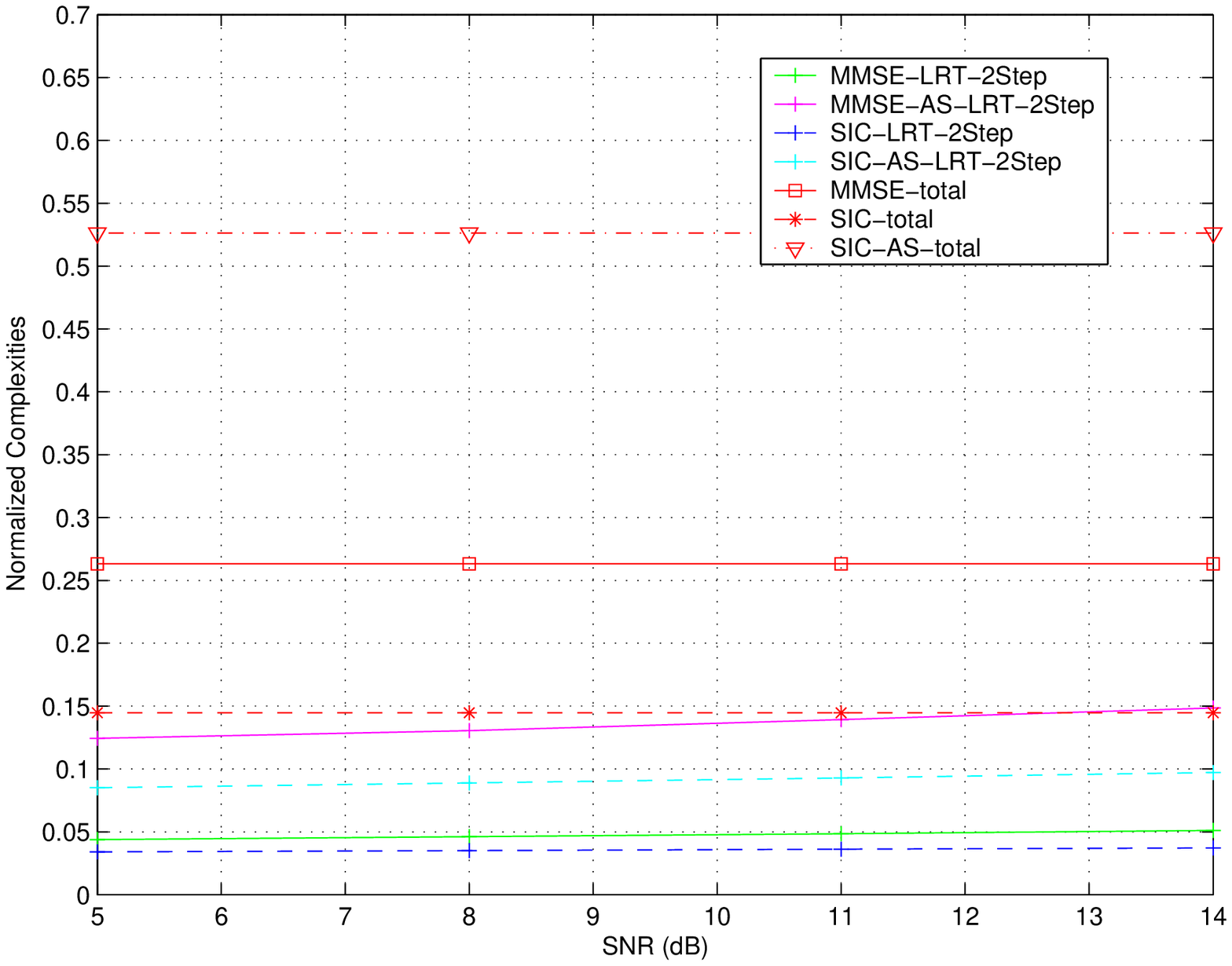}
	\caption{Normalized total metric computation complexity versus SNR (dB)}
	\label{fig_plotncomp}
	\end{minipage}
	\hspace{0.5cm}
	\begin{minipage}[b]{0.45\linewidth}
	\centering
	\includegraphics[width=\textwidth]{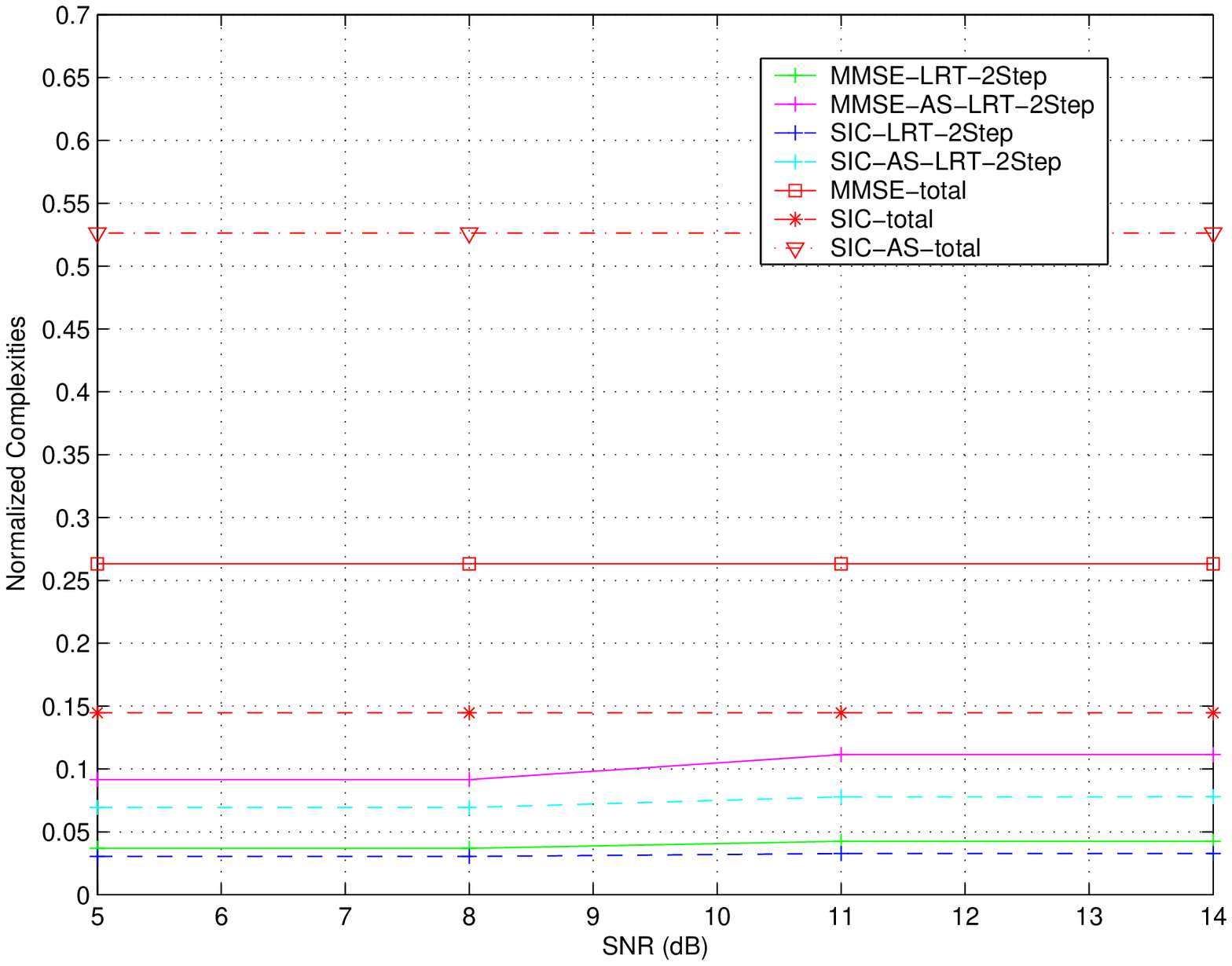}
	\caption{Normalized total metric computation complexity versus SNR (dB): Computation complexity of a metric scales with chunk length.}
	\label{fig_plotncompCL}
	\end{minipage}
	\end{figure}

     In Figures \ref{fig_plotscmmse} to \ref{fig_plotncomp} we assume that $N= 20$ RBs are available   for serving $K=10$ active users, all of whom have identical maximum transmit powers.
        In Fig. \ref{fig_plotscmmse}, we plot the average cell spectral efficiency (in bits-per-sec-per-Hz) versus the average transmit SNR (dB) for an uplink where each user has one transmit antenna and the BS employs the linear MMSE receiver. We plot the spectral efficiencies achieved when Algorithm IIa is employed with and without the second phase (described in Section \ref{sec:SP}), respectively (denoted in the legend by MU-MMSE-LRT-2Step and MU-MMSE-LRT-1Step). Also plotted is the upper bound obtained by the linear programming (LP) relaxation of (P1) along with the spectral efficiency obtained upon rounding the LP solution to ensure feasibility (denoted in the legend by MU-MMSE-LP-UB and MU-MMSE-LP-Rounding, respectively).

        In Fig. \ref{fig_plotscsic}, we plot the average cell spectral efficiency   versus the average transmit SNR   for an uplink where each user has one transmit antenna and the BS employs the successive interference cancelation (SIC) receiver. We plot the spectral efficiencies achieved when Algorithm IIa is employed with and without the second phase, respectively (denoted in the legend by MU-SIC-LRT-2Step and MU-SIC-LRT-1Step). Also plotted are the corresponding LP upper bound along with the spectral efficiency obtained upon rounding the LP solution. The counterparts of Figures \ref{fig_plotscmmse} and   \ref{fig_plotscsic} in the scenario where each user has two transmit antennas and the BS can thus exploit transmit antenna selection are given in  Figures \ref{fig_plotscmmseas} and \ref{fig_plotscsicas}, respectively.

        Finally, in Fig. \ref{fig_plotnse}   we plot the normalized spectral efficiencies obtained by dividing each spectral efficiency by the one yielded by Algorithm IIa when only single user (SU) scheduling is allowed, which in turn can be emulated by setting all metrics  $p(\Uc,\cb): (\Uc,\cb)\in\Mk$ in (P1) to be zero whenever $|\Uc|\geq 2$.\footnote{Note that for SU scheduling MMSE and SIC receivers are equivalent.} In all considered schemes we assume that Algorithm IIa with the second phase is employed.
     From Figures \ref{fig_plotscmmse} to \ref{fig_plotnse}, we have the following observations:
   \begin{itemize}
   \item For both SIC and MMSE receivers, the performance of Algorithm IIa is more than  $80 \%$ of the respective LP upper bounds, which is much superior to the worst case guarantee $1/3$ (obtained by specializing the result in  (\ref{eq:pRAlte1}) by setting $\Mk^{\rm wide}=\phi,T=2$ and $\Delta=J=0$). Further, for both the receivers the performance of Algorithm IIa with the second phase is more than  $90\%$ the respective LP upper bounds. The same conclusions can be drawn when antenna selection is also exploited by the BS.  In all cases, the performance of LP plus rounding scheme is exceptional and within $2 \%$ of the respective   upper bound. However the complexity of this LP seems unaffordable as yet for practical implementation.\footnote{For instance, this LP involves about $11,500$ variables and must be solved within each scheduling interval whose duration in LTE systems is one millisecond.}
       \item The SIC receiver results in a small gain  ($1.5\%$ to $2.5\%$) over the MMSE receiver. This gain will increase if we consider more correlated fading over which the limitation of linear receivers is exposed and as the maximum number of users that can be co-scheduled on an RB ($T$) is increased since the SIC allows for  improved system rates via co-scheduling a larger number of users on an RB, whereas the MMSE will become interference limited. Note that antenna selection seems to provide a much larger gain ($6\%$ to $8\%$) that the one offered by the advanced SIC receiver. This observation must be tempered by the facts that the simulated scenario of independent (uncorrelated) fading is favorable for antenna selection and that the antenna switching loss (about $0.5$ dB in practical devices) as well as the additional pilot overhead have been neglected. 
       \item MU scheduling offers  substantial gains over SU scheduling (ranging from $50\%$ to $75\%$ for the considered SNRs). This follows since the degrees of freedom available here for MU scheduling is twice that of SU-scheduling.
   \end{itemize}

   Next, in Fig. \ref{fig_plotncomp} we plot the normalized total metric computation complexities for the scheduling schemes considered in Figures \ref{fig_plotscmmse} to \ref{fig_plotnse}. 
   In all cases the second phase is performed for Algorithm IIa and more importantly the sub-additivity property together with the on-demand metric computation feature are exploited, as described in Section \ref{sec:compred}, to avoid redundant metric computations. All schemes compute the SU metrics $\{p(\Uc,\cb):(\Uc,\cb)\in\Mk\;\&\;|\Uc|=1\}$. The cost assumed for computing each metric is given in Table \ref{tab:defmetc}.  
   Note that the cost of an MU metric for the SIC receiver is smaller because with this receiver one of the users sees an interference free channel. Thus, its contribution to the metric is equal to the already computed SU metric determined for the allocation when that user is scheduled alone on the corresponding chunk, and hence need not be counted in the cost.

   We use MMSE-Total and SIC-Total to denote the total metric computation complexities obtained with the MMSE receiver and the SIC receiver, respectively,  by counting the corresponding complexities for all pairs $(\Uc,\cb)\in\Mk$, whereas   MMSE-AS-Total, SIC-AS-Total denote their counterparts when antenna selection is also exploited by the BS. Note that all complexities in Fig. \ref{fig_plotncomp} are normalized by MMSE-AS-Total. The key takeaway from Fig. \ref{fig_plotncomp} is that exploiting sub-additivity together with the on-demand metric computation can result in very significant   metric computation complexity reduction. In particular, in this example more than $80\%$   reduction is obtained for the MMSE receiver  and more than $75\%$ reduction is obtained for the SIC receiver,  with the respective gains being larger when antenna selection is also exploited. 
     Further, we note that considering Algorithm IIa, the second phase itself adds a very small   metric computation complexity overhead but results in a large performance improvement. To illustrate this, for the MMSE receiver the complexity overhead ranges from $2$ to $4\%$, whereas the performance improvement ranges from $9$ to $13\%$, respectively. Then, in Fig. \ref{fig_plotncompCL} we consider the same setup as in Fig. \ref{fig_plotncomp} but now the computational complexity of each $p(\Uc,\cb)$ also scales with the length of the chunk indicated by $\cb$. From Fig. \ref{fig_plotncompCL} we see that the   metric computation complexity reductions are even larger.

\begin{figure}[ht]
	\begin{minipage}[b]{0.45\linewidth}
	\centering
	\includegraphics[width=\textwidth]{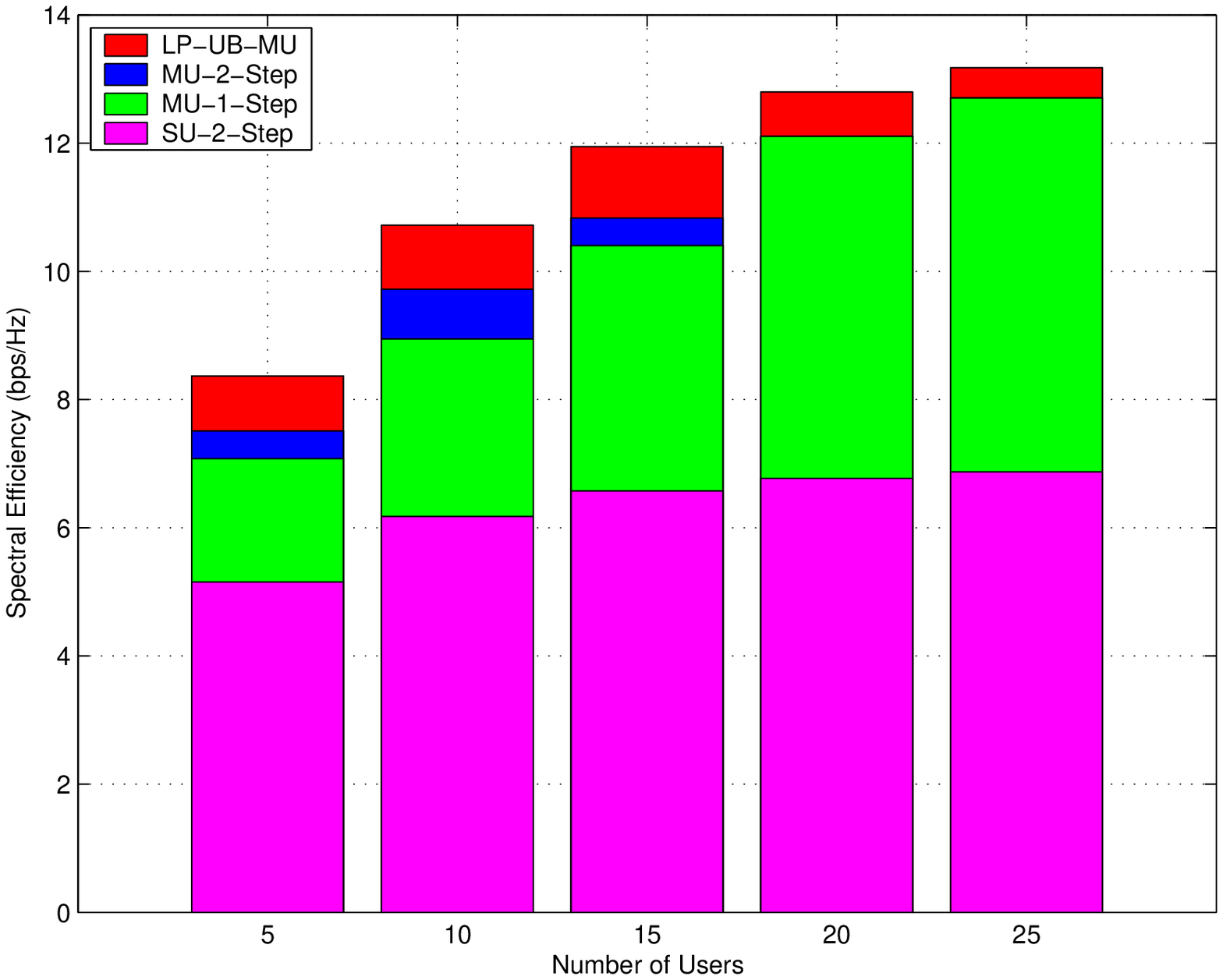}
	\caption{Average spectral efficiency versus number of users: 5dB SNR}
	\label{fig:figure5db}
	\end{minipage}
	\hspace{0.5cm}
	\begin{minipage}[b]{0.45\linewidth}
	\centering
	\includegraphics[width=\textwidth]{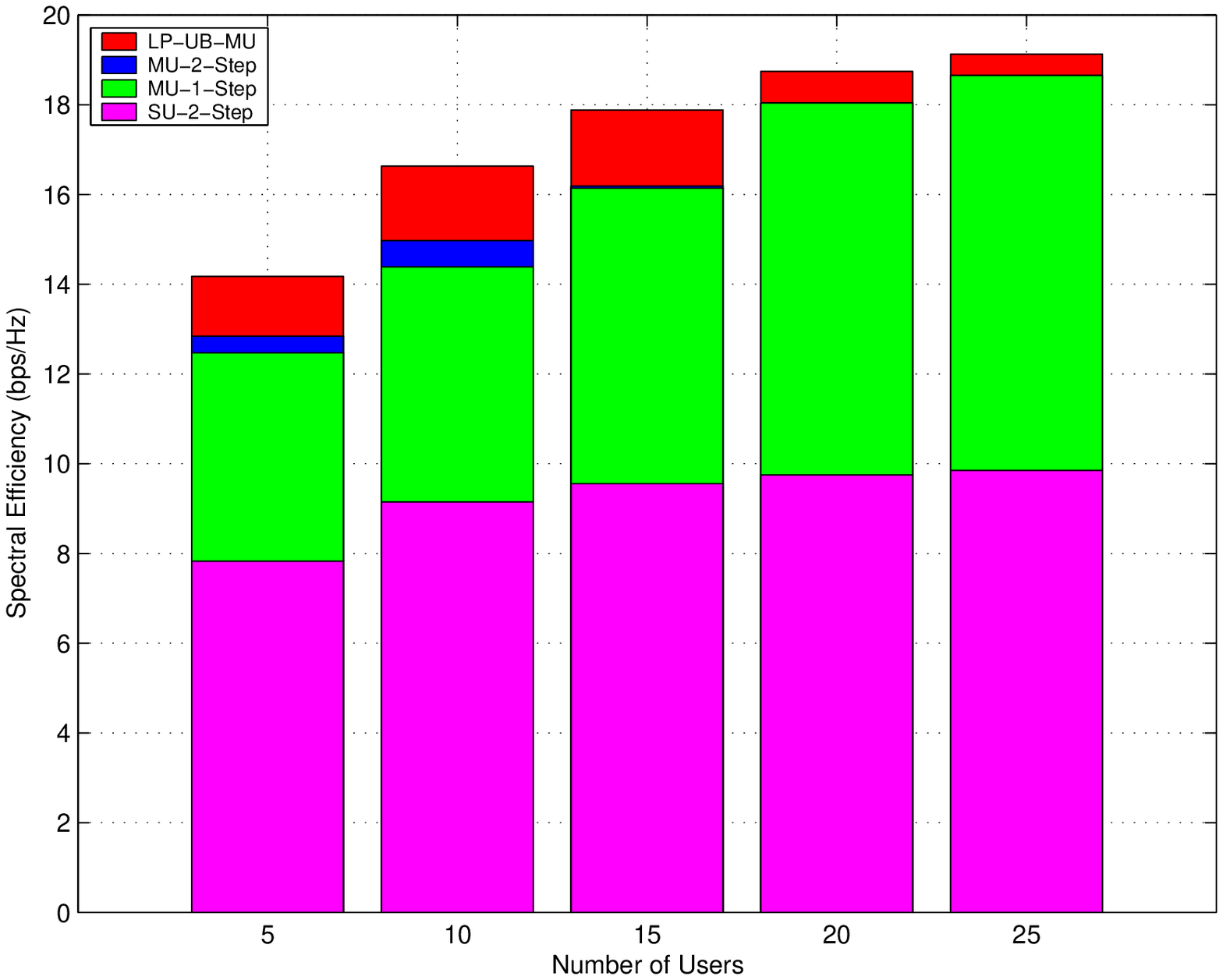}
	\caption{Average spectral efficiency versus number of users: 14dB SNR}
	\label{fig:figure14db}
	\end{minipage}
	\end{figure}

Finally, in Figures \ref{fig:figure5db} and \ref{fig:figure14db} we consider an UL with $N=10$ RBs and where each user has one transmit antenna while the BS employs the linear MMSE receiver. We  plot the average cell spectral efficiency versus the number of users  for a given transmit SNR.  From the plots  we see that MU scheduling maintains a significant gain over SU scheduling. Interestingly, the gain of the second phase on Algorithm IIa in MU scheduling reduces as the number of users exceeds the number of RBs, whereas the solution yielded by Algorithm IIa (without the second phase) approaches the optimal one since the gap to the LP upper bound vanishes.

\begin{table}
\begin{small}
\begin{center}
\hspace{-1cm}\caption[]{Metric computation complexity}\label{tab:defmetc}
\begin{tabular}{|p{5cm}|p{5cm}|p{5cm}|}\hline
   Metric & MMSE &  SIC \\ \hline
   $p(\Uc,\cb):|\Uc|=1\;\&\;N_t=1$ & 1 & 1 \\ \hline   $p(\Uc,\cb):|\Uc|=1,\; N_t=2\;\&\;{\rm Ant.Sel.}$ & 2 & 2   \\ \hline
 $p(\Uc,\cb):|\Uc|=2\;\&\;N_t=1$ & 2  & 1 \\ \hline   $p(\Uc,\cb):|\Uc|=2,\; N_t=2\;\&\;{\rm Ant.Sel.}$ & 8 & 4\\ \hline
     \end{tabular}
\end{center}
\end{small}
\end{table}

%

\section{Sequential LRT based MU Scheduling}\label{sec:seq}
We next propose a sequential LRT based MU scheduling method that yields a scheduling decision over $\Mk^{\rm narrow}$ . As before, our focus is on avoiding as many metric computations as possible. The idea is to implement the LRT based MU scheduling algorithm in $T$ iterations, where we recall $T$ denotes the maximum number of users that can be co-scheduled on an RB. In particular, in the first iteration we define metrics $\breve{p}(\Uc,\cb)=p(\Uc,\cb),\;\forall\;(\Uc,\cb)\in \Mk^{\rm narrow}:|\Uc|=1$ with $\breve{p}(\Uc,\cb)=0$ otherwise, and
use these metrics in Algorithm IIa to obtain a tentative scheduling decision. Further,
 in the $s^{th}$ iteration where $2\leq s\leq T-1$, we first perform the following steps to obtain metrics $\breve{p}(\Uc,\cb),\;\forall\;(\Uc,\cb)\in \Mk^{\rm narrow}$, where only a few of these metrics are positive, and then use them in Algorithm IIa to obtain a tentative   decision.
\begin{itemize}
\item Initialize $\breve{p}(\Uc,\cb)=0,\;\forall\;(\Uc,\cb)\in \Mk^{\rm narrow}$. Let $\Sc'$ denote the output obtained from the previous iteration.
    \item For each $(\Uc,\cb)\in\Sc'$ we ensure that any user in set $\Uc$ can be scheduled in the $s^{th}$ iteration only as part of a set that contains all users in $\Uc$ along with at-most one additional user, by setting
    \begin{eqnarray*}
   \breve{p}(\Uc',\cb') = 0 \;{\rm if}\; (\Uc\not\subseteq\Uc'\;\&\;\Uc'\cap\Uc\neq\phi)\;{\rm or}(|\Uc'|>|\Uc|+1\;\&\;\Uc'\cap\Uc\neq\phi),\;\;\forall\;(\Uc',\cb')\in \Mk^{\rm narrow}.
 \end{eqnarray*}
 \item For each $(\Uc,\cb)\in\Sc'$, we also ensure that any user in set $\Uc$ must be assigned all RBs in $\cb$, by considering each $(\Uc',\cb')\in \Mk^{\rm narrow}:\Uc\subseteq\Uc'\;\&\; |\Uc'|\leq |\Uc|+1$, and setting
\begin{eqnarray*}
    \breve{p}(\Uc',\cb') &=& \left\{\begin{array}{c} p(\Uc',\cb'),\;\; {\rm If}\;{\rm Tail(\cb')}\geq {\rm Tail(\cb)} \;\&\; {\rm Head(\cb')}\leq {\rm Head(\cb)}\\
  0,\;\;{\rm Otherwise}.\end{array}\right.\\
 \end{eqnarray*}
\end{itemize}
In the last iteration, i.e. when $s=T$, we  initialize $\breve{p}(\Uc,\cb)=p(\Uc,\cb),\;\forall\;(\Uc,\cb)\in \Mk^{\rm narrow}$. Then, using the set $\Sc'$ obtained as the output  of the $(T-1)^{th}$ iteration, we perform the  two aforementioned steps. Additionally, to ensure non-overlapping chunk allocation, for each $(\Uc,\cb)\in\Sc'$ we set
\begin{eqnarray*}
   \breve{p}(\Uc',\cb') = 0 \;{\rm if}\; \cb'\cap\cb\neq\phi \;\&\;\Uc'\cap\Uc=\phi,\;\forall\;(\Uc',\cb')\in \Mk^{\rm narrow}.
 \end{eqnarray*}
 Note that the  different initialization chosen for the last iteration seeks to   select  a larger pool of positive  metrics and can improve performance albeit at an increased complexity. In addition, after each iteration we also enforce an {\em improvement condition} which checks if the weighted sum rate yielded by the obtained decision is strictly greater than that computed at the end of the previous iteration. If this condition is satisfied, we proceed to the next iteration, else the process is terminated and the solution obtained at the end of the previous iteration is returned.
Notice that in each iteration  only a small subset out of the set of all metrics is selected, which in particular is that whose  corresponding pairs are compatible (as defined in the aforementioned conditions) with the output tentative scheduling decision of the previous iteration. 
Next, we offer an approximation result for the sequential LRT based MU scheduling that holds under mild assumptions.
\begin{assumption}\label{assump2}
Suppose $\Fc$ is any allocation $\{(\Uc,\cb)\}$ that is feasible for (P1). Then $\Fc$ is downward closed in the following sense. 
Any allocation $\Fc'$ constructed as $\Fc'=\{(\Uc',\cb):\Uc'\subseteq\Uc\;\&\;(\Uc,\cb)\in\Fc\}$ is also feasible.
\end{assumption}
\begin{theorem}
Suppose that Assumptions \ref{assump1} and \ref{assump2} are satisfied.  Let the weighted sum rate yielded by the sequential LRT based MU scheduling over $\Mk^{\rm narrow}$ be denoted by $\hat{W}^{\rm seq-narrow}$. Then,
\begin{eqnarray}\label{eq:pRAlteSeq}
  \hat{W}^{\rm seq-narrow}\geq
  \frac{\hat{W}^{\rm opt,narrow}}{T(2+\Delta+2J)}.
 \end{eqnarray}
\end{theorem}
\proof
Let $\Fc ^{\rm opt,narrow}$ be an optimal allocation of pairs from $\Mk^{\rm narrow}$ that yields   a weighted sum rate $\hat{W}^{\rm opt,narrow}$ and initialize $\Fc'=\phi$. Then for each $(\Uc,\cb)\in\Fc ^{\rm opt,narrow}$ determine the best user $\hat{u}=\arg\max_{u\in\Uc}\{p(u,\cb)\}$ and insert the pair $(\hat{u},\cb)$ into $\Fc'$. Note that due the sub-additivity property in Assumption  \ref{assump1}, we must have
 that $p(\hat{u},\cb)\geq\frac{p(\Uc,\cb)}{T}$. Consequently, we have that the weighted sum rate yielded by $\Fc'$ is at-least $\frac{\hat{W}^{\rm opt,narrow}}{T}$.
 Furthermore, on account of Assumption \ref{assump2}, $\Fc'$ is a feasible allocation for (P1). Then, suppose $\Fc^{(1)}$ is the allocation obtained after the first iteration of the sequential algorithm. Since this allocation is a result of applying Algorithm IIa with single user metrics, upon invoking Theorem 1 we can claim that the weighted sum rate yielded by $\Fc^{(1)}$ is at-least a fraction $\frac{1}{2+\Delta+2J}$ of the best single-user allocation, where a single-user allocation is one where each pair includes only one user. Then, since $\Fc'$ is one such single-user allocation we can claim that
  the weighted sum rate yielded by $\Fc^{(1)}$ is at-least    $\frac{\hat{W}^{\rm opt,narrow}}{T}$. Finally, since the improvement condition ensures that the weighted sum rates yielded by tentative allocations across iterations are monotonically increasing , we can deduce that the theorem is true. \endproof

\section{User Pre-Selection}\label{sec:usel}
In a practical cellular system the number of active users can be large. Indeed the control channel constraints may limit the BS to serve a much smaller subset of users. It thus makes sense from a complexity stand-point to pre-select a pool of {\em good} users and then use the MU scheduling algorithm on the selected pool of users.
Here we propose a few user pre-selection algorithms. For convenience, wherever needed, we assume that at-most two users can be co-scheduled on an RB (i.e., $T=2$) which happens to be the most typical value.

Before proceeding we need to define some terms that will be required later. Suppose that each user has one transmit antenna and let $\hb_{u,j}$ denote the effective channel vector seen at the BS from user $u$ on RB $j$, where $1\leq u\leq K$ and $1\leq j\leq N$. Note that the effective channel vector includes the fading as well as the path loss factor and a transmit power value. Then, letting  $\alpha_u$ denote the  PF weight of user $u$, we define the following metrics:
\begin{itemize}
\item Consider first the weighted rate that the system  can obtain when it   schedules user $u$ alone on   RB $j$,
 \begin{eqnarray}
  p^{\rm su}(u,j)&=& \alpha_u\log(1 + \hb_{u,j}^{\dag}\hb_{u,j}),\;\;\forall\;1\leq u\leq K\;\&\;1\leq j\leq N.
 \end{eqnarray}
\item Let $\Uc=\{u,v\}:u\neq v$ be any pair of users   and suppose that the BS employs the MMSE  receiver. Then, the weighted sum rate obtained by scheduling the user pair $\Uc$ on RB $j$ is given by
    \begin{eqnarray}
  p^{\rm mmse}(\Uc,j)&=& \alpha_u\log(1 + \hb_{u,j}^{\dag}(\Ib + \hb_{v,j}\hb_{v,j}^{\dag})^{-1}\hb_{u,j})+ \alpha_v\log(1 + \hb_{v,j}^{\dag}(\Ib + \hb_{u,j}\hb_{u,j}^{\dag})^{-1}\hb_{v,j}).
 \end{eqnarray}
\item Finally, assume that the BS employs the SIC  receiver and let $\hat{u}=\arg\max_{s\in\Uc}\{\alpha_s\}$ and let $\hat{v}=\Uc\setminus\hat{u}$.
Then, the weighted sum rate obtained by scheduling the user pair $\Uc$ on RB $j$ is given by
    \begin{eqnarray}\label{eq:sicutil}
  p^{\rm sic}(\Uc,j)&=& p^{\rm su}(\hat{u},j)+ \alpha_{\hat{v}}\log(1 + \hb_{\hat{v},j}^{\dag}(\Ib + \hb_{\hat{u},j}\hb_{\hat{u},j}^{\dag})^{-1}\hb_{\hat{v},j}).
 \end{eqnarray}
 \end{itemize}

 We are now ready to offer our user pre-selection rules where a pool of $\tilde{K}$ users must be selected from the $K$ active users. Notice that to reduce complexity, all rules neglect the contiguity and the complete overlap constraints.
 \begin{enumerate}
\item The first rule simply selects the $\tilde{K}$ users that offer the $\tilde{K}$ largest single-user rates among $\{\sum_{j=1}^Np^{\rm su}(u,j)\}_{u=1}^K$.
 \item The second rule assumes that each RB can be assigned to at-most one user. Then, if a user subset $\Ac\subseteq\{1,\cdots,K\}$ is selected, the system weighted sum-rate is given by
     \begin{eqnarray}
   f(\Ac)\define \sum_{j=1}^N\max_{u\in\Ac}\{p^{\rm su}(u,j)\}.
 \end{eqnarray}
     It can be shown that $f:2^{\{1,\cdots,K\}}\to\Reals_+$ is a {\em monotonic sub-modular set function} \cite{zhangP:2012}.
     As a result, the user pre-selection problem
      \begin{eqnarray}
   \arg\max_{ \Ac\subseteq\{1,\cdots,K\}:|\Ac|\leq \tilde{K}}\{f(\Ac)\}
 \end{eqnarray}
can be sub-optimally solved by adapting a simple greedy algorithm \cite{nemhaus:algo}, which offers a half approximation \cite{zhangP:2012}.
\item The third rule assumes that each RB can be assigned to at-most two users and that the BS employs the MMSE receiver. Then, if a user subset $\Ac\subseteq\{1,\cdots,K\}$ is selected, the system weighted sum-rate is given by
     \begin{eqnarray}\label{eq:metpairmmse}
   g(\Ac)\define \sum_{j=1}^N\max\left\{\max_{u\in\Ac}\{p^{\rm su}(u,j)\},\max_{\Uc=(u,v):u<v\atop u,v\in\Ac}\{p^{\rm mmse}(\Uc,j) \}\right\}.
 \end{eqnarray}
     It can be shown that $g:2^{\{1,\cdots,K\}}\to\Reals_+$ is a monotonic set function but unfortunately it need not be sub-modular. Nevertheless, we  proceed to employ the greedy algorithm to sub-optimally solve
      \begin{eqnarray}
   \arg\max_{ \Ac\subseteq\{1,\cdots,K\}:|\Ac|\leq \tilde{K}}\{g(\Ac)\}
 \end{eqnarray}
      \item The fourth rule also assumes that each RB can be assigned to at-most two users but that the BS employs the SIC receiver. However, even upon replacing $p^{\rm mmse}(\Uc,j)$ in (\ref{eq:metpairmmse}) with $p^{\rm sic}(\Uc,j)$, the resulting set function need not be sub-modular. As a result we use a different metric. In particular,  for a user subset $\Ac\subseteq\{1,\cdots,K\}$ we employ a metric that is given by
     \begin{eqnarray}\label{eq:metpairsic}
\nonumber   h(\Ac)\define \sum_{j=1}^N   \sum_{\Uc=(u,v):u\leq v\atop u,v \in\{1,\cdots,K\}}\left(p^{\rm su}(\Uc\cap\Ac,j)\Xc(|\Uc\cap\Ac|=1)+ p^{\rm sic}(\Uc,j)\Xc(|\Uc\cap\Ac|=2)\right) \\
   =\sum_{j=1}^N\left((K-|\Ac|+1)\sum_{u\in\Ac}p^{\rm su}(u,j) +  \sum_{\Uc=(u,v):u<v\atop u,v \in\Ac}p^{\rm sic}(\Uc,j) \right).
 \end{eqnarray}
 Notice that for any $\Ac$, $h(\Ac)$ represents the system weighted sum-rate when  time-sharing is employed by the system wherein in each slot only a particular user or two distinct users from a particular pair in  $\{1,\cdots,K\}$ are allowed to be scheduled. Then, a key result 
 is the following.
 \begin{theorem}\label{prop:sbm}
 The set function $h(.)$ defined in (\ref{eq:metpairsic}) is a  monotonic sub-modular set function. Thus the problem
   \begin{eqnarray}
   \arg\max_{ \Ac\subseteq\{1,\cdots,K\}:|\Ac|\leq \tilde{K}}\{h(\Ac)\}
 \end{eqnarray}
 can be solved sub-optimally (with a $1/2$ approximation) by a simple greedy algorithm.
\end{theorem}
\proof On any RB $j$, consider any fixed pair $\Uc=\{u,v\}\subseteq\{1,\cdots,K\}$ and define the set function
 \begin{eqnarray}
 g(\Ac)=p^{\rm su}(\Uc\cap\Ac,j)\Xc(|\Uc\cap\Ac|=1)+ p^{\rm sic}(\Uc,j)\Xc(|\Uc\cap\Ac|=2),\;\forall\;\Ac\subseteq\{1,\cdots,K\}.
 \end{eqnarray}
Our first aim is to prove that $g(.)$ defined above is a monotonic sub-modular set function. First, note that the weighted sum rate in (\ref{eq:sicutil}) can also be written as,
\begin{eqnarray}\label{eq:sicutil2}
  \nonumber p^{\rm sic}(\Uc,j)&=& (\alpha_{\hat{u}}-\alpha_{\hat{v}})p^{\rm su}(\hat{u},j)+ \alpha_{\hat{v}}\log\left|\Ib + \hb_{\hat{v},j}\hb_{\hat{v},j}^{\dag}+ \hb_{\hat{u},j}\hb_{\hat{u},j}^{\dag}\right|\\
  &\geq& p^{\rm su}(\hat{v},j)+ \alpha_{\hat{u}}\log(1 + \hb_{\hat{u},j}^{\dag}(\Ib + \hb_{\hat{v},j}\hb_{\hat{v},j}^{\dag})^{-1}\hb_{\hat{u},j})
 \end{eqnarray}
so that $p^{\rm sic}(\Uc,j)\geq\max\{p^{\rm su}(\hat{u},j),p^{\rm su}(\hat{v},j)\}$, which suffices to prove the monotonicity of $g(.)$. Then, to prove sub-modularity we must show that,
\begin{eqnarray}\label{eq:sbutil}
 g(\Ac\cup\{q\})-g(\Ac)\geq g(\Bc\cup\{q\})-g(\Bc),\;\forall\;\Ac\subseteq\Bc\subseteq\{1,\cdots,K\}\;\&\;q\in\{1,\cdots,K\}\setminus\Bc.
 \end{eqnarray}
To prove (\ref{eq:sbutil}) we consider any $\Ac\subseteq\Bc\subseteq\{1,\cdots,K\}$ so that $\Ac\cap\Uc\;\subseteq\;\Bc\cap\Uc$ and  consider the following cases. First consider the case, $|\Ac\cap\Uc|=|\Bc\cap\Uc|$ which implies that both $\Ac,\Bc$ contain the same user(s) from $\Uc$ so that (\ref{eq:sbutil}) must hold with equality.
Then, suppose $|\Ac\cap\Uc|<|\Bc\cap\Uc|$. In this case, upon exploiting the inequality
\begin{eqnarray}\label{eq:sicutil3}
  p^{\rm sic}(\Uc,j)\leq  p^{\rm su}(u,j)+  p^{\rm su}(v,j),
 \end{eqnarray}
 together with the fact that $g(\Bc\cup\{q\})-g(\Bc)=0$ when $|\Bc\cap\Uc|=2$,
 we can conclude that (\ref{eq:sbutil}) must hold. Then, since the set function $h(.)$ in (\ref{eq:metpairsic}) is a linear combination of $NK(K+1)/2$ monotonic sub-modular set functions in which the combining coefficients are all positive, we can assert that it must be a monotonic sub-modular set function as well.\endproof
 \end{enumerate}

As a benchmark to compare the performance of the proposed user pre-selection algorithms we can consider the case where LRT MU scheduling is employed without user pre-selection but where an additional knapsack constraint is used to enforce the limit on the number of users that can be scheduled in an interval. It can be verified that this can be achieved by defining a knapsack constraint in (P1) as $\beta^1(\Uc,\cb)=\frac{|\Uc|}{\tilde{K}},\;\forall\;(\Uc,\cb)\in\Mk$. 

\section{System Level Simulation Results}
We now present the performance of our MU scheduling algorithms (including the sequential algorithm of Section \ref{sec:seq} and the user pre-selection schemes of Section \ref{sec:usel}) via detailed system level simulations which were conducted on a fully calibrated system simulator that we developed. The simulation parameters conform to those used in 3GPP LTE evaluations and are given in Table~\ref{tab:parameter}. In all cases inter-cell interference suppression (IRC) is employed by each base-station (BS).

We first consider the case when each cell (or sector) has an average of 10 users and where there are no knapsack constraints. In Table \ref{tab:resSUMU} we report the cell average and cell edge spectral efficiencies. The percentage gains shown for the MU scheduling schemes are over the baseline LRT based single-user scheduling scheme. Note that for the first three scheduling schemes we employed the second phase described in Section \ref{sec:SP}. Also, we observed that the LRT based SU scheduling together with the second phase yields at-least as good a  performance (for both cell-edge and cell average throughputs) as those of the deterministic SU scheduling algorithms in \cite{multiserver:2009,Lee-UL-2009}, so we have omitted results for the latter algorithms. As seen from  Table \ref{tab:resSUMU}, MU scheduling in conjunction with an advanced SIC receiver at the BS can result in very significant gains in terms of cell average throughout (about $27\%$) along with good cell edge gains. For the simpler MMSE receiver, we see significant   cell average throughout gains (about $18\%$) but a degraded cell edge performance.
Finally, the last two reported schemes are based on the sequential-LRT method described in Section  \ref{sec:seq}. We notice that  sequential-LRT based scheduling provides significant cell average gains while retaining the cell edge performance of SU scheduling. Thus, the sequential LRT based scheduling method is an attractive way to tradeoff some cell average throughput gains for a reduction in complexity.

Next, in Tables \ref{tab:resMUselmmse} and \ref{tab:resMUselsic} we consider LRT based MU scheduling, with the second phase described in Section \ref{sec:SP}, for the case when the BS employs the MMSE receiver and the case when it employs the SIC receiver, respectively. In each case we assume that an average of 15 users are present in each cell and at-most 7 first-transmission users can be scheduled in each interval. Thus, a limit on the number of scheduled users might have to  be enforced in each scheduling interval. As a benchmark, we enforce this constraint (if it is required) using one knapsack constraint as described in Section \ref{sec:usel}. Note that upon specializing the result in Theorem 1 (with  $\Mk^{\rm wide}=\phi,T=2$ and $\Delta=0,J=1$)) we see that the LRT based MU scheduling algorithm  guarantees an approximation factor of $1/5$.
Then, we examine the scenario where a pool of $\tilde{K}=7$  users is pre-selected whenever the number of
first-transmission users is larger than $7$. The LRT based MU scheduling algorithm is then employed on this pool without any constraints. In Table \ref{tab:resMUselmmse} we have used the first second and third pre-selection rules from Section \ref{sec:usel} whereas in Table \ref{tab:resMUselsic} we have used the first second and fourth pre-selection rules.
It is seen that the simple rule one provides a superior performance compared to the benchmark. Indeed, it is attractive since it involves computation of only single user metrics. The other rule (rule 2) which possess this feature, however provides much less improvement mainly because it is much more aligned to single user scheduling. Rules 3 and 4 involve computation of metrics that involve user-pairing and hence incur higher complexity. For the MMSE receiver, the gain of rule 3 over rule 1 is marginal mainly because the metric in rule 3 is not sub-modular and hence cannot be well optimized by the simple greedy rule. On the other hand, considering the MMSE receiver, the gain of rule 4 over rule 1 is larger because the metric used in rule 4 is indeed sub-modular and hence can  be well optimized by the simple greedy rule.

\begin{table}[t]
   \centering
 \begin{tabular}{|l|l|}
    \hline
    \textbf{Parameter}  & \textbf{Assumption}\\\hline
Deployment scenario & IMT Urban Micro (UMi)\\\hline
Duplex method and bandwidth & FDD: 10MHz for uplink\\\hline
Cell layout & Hex grid 19 sites, 3 cells/site\\\hline
Transmission power at user &  23 dBm \\\hline
Average number of users per sector  & 10 or 15\\\hline
Network synchronization & Synchronized\\\hline
Antenna configuration (eNB) & 4 RX co-polarized ant., 0.5-$\lambda$ spacing \\\hline
Antenna configuration (user)    & 1 TX ant. \\\hline
Uplink transmission scheme    & Dynamic MU scheduling,  \\& MU pairing: Max 2/RB  users aligned pairing;\\\hline
Fairness metric    & Proportional Fairness\\\hline
Fractional power control  & Po=-85 dB, $\alpha=0.8$  \\\hline
Uplink scheduler  & PF in time and frequency\\\hline
Scheduling granularity: &  1 RB\\\hline
Uplink HARQ scheme &  Synchronous, non-adaptive Chase Combining\\\hline
Uplink receiver type &    MMSE-IRC and SIC-IRC \\\hline
Channel estimation error &  NA\\\hline
 \end{tabular}
  \caption{\small  Parameters for system level simulations}\label{tab:parameter} \vspace{1em}
\end{table}

\begin{table}[t]
   \centering
 \begin{tabular}{|l|l|l|}
    \hline
    Scheduling method & cell average     & 5\% cell-edge  \\ \hline
    LRT SU & 1.6214 & 0.0655\\ \hline
    LRT MU with MMSE   & 1.9246 (18.70\%) & 0.0524 \\\hline
    LRT MU with SIC   & 2.0651 (27.37\%)  & 0.0745 \\\hline
    LRT-Sequential MU with MMSE   & 1.8196 (12.22\%) & 0.0627 \\\hline
    LRT-Sequential MU with SIC   & 1.9537 (20.5\%)  & 0.0665 \\\hline
              \end{tabular}
  \caption{\small Spectral efficiency  of LRT based SU and MU UL scheduling schemes.  An average of 10 users are present in each cell and all associated active users can be scheduled in each interval.}\label{tab:resSUMU} \vspace{1em}
\end{table}

\begin{table}[t]
   \centering
 \begin{tabular}{|l|l|l|}
    \hline
    LRT-MU scheduling with: & cell average     & 5\% cell-edge  \\ \hline
    Knapsack constraint & 1.7833 & 0.0266\\ \hline
    pre-selection 1    & 1.7940 (0.6\%)  & 0.0419 (57.52\%)\\\hline
    pre-selection 2   & 1.7908 (0.4\%)  & 0.0414 (55.64\%)\\ \hline
    pre-selection 3 & 1.8265 (2.42\%) &  0.0444 (66.92\%) \\ \hline
              \end{tabular}
  \caption{\small Spectral efficiency  of MU UL scheduling schemes with MMSE receiver.  An average of 15 users are present in each cell and at-most 7 first-transmission users can be scheduled in each interval.}\label{tab:resMUselmmse} \vspace{1em}
\end{table}

\begin{table}[t]
   \centering
 \begin{tabular}{|l|l|l|}
    \hline
    LRT-MU scheduling with: & cell average     & 5\% cell-edge  \\ \hline
    Knapsack constraint & 1.8865 & 0.0411\\ \hline
    pre-selection 1    & 2.0082 (6.45\%) & 0.0527 (28.22\%)\\\hline
    pre-selection 2   & 1.8980 (0.61\%)  & 0.0451 (9.73\%)\\ \hline
    pre-selection 4 & 2.1069 (11.68\%) &  0.0531 (29.2\%) \\ \hline
              \end{tabular}
  \caption{\small Spectral efficiency  of MU UL scheduling schemes with SIC receiver.  An average of 15 users are present in each cell and at-most 7 first-transmission users can be scheduled in each interval.}\label{tab:resMUselsic} \vspace{1em}
\end{table}

\section{Conclusions and Future Research}
We considered resource allocation in the   3GPP LTE  cellular uplink  wherein multiple users can be assigned the same time-frequency resource.  We showed that the resulting resource allocation problem, which must comply with several practical constraints, is  APX-hard. We then   proposed constant-factor polynomial-time   approximation  algorithms   and demonstrated their performance via simulations. An interesting avenue for future work is to obtain  good bounds on the average case performance of our proposed algorithms. In addition, the design of a joint scheduling algorithm that also determines assignment of control channel resources to the active users is an important open problem.


\section{Appendix: Modeling 3GPP LTE Control Channel Constraints}
Note that by placing restrictions on the location where a particular user's control packet can be sent and the size of that packet, the system can  reduce the number of blind decoding attempts that have to be made by that user in order to receive its control packet. We note that a user is unaware of whether  there is a control packet intended for it and consequently must check all possible locations where its control packet could be present assuming each possible packet size.  Each control packet carries a CRC bit sequence scrambled using the unique user identifier which helps the user deduce whether the examined packet is meant for it. 
In the 3GPP LTE system, the minimum allocation unit in the downlink control channel  is referred to as the control channel element (CCE). Let $\{1,\cdots,R\}$ be a set of  CCEs available for conveying UL grants. A contiguous chunk of CCEs from $\{1,\cdots,R\}$ that can be   be assigned to a   user is referred to as a PDCCH. The size of each PDCCH is referred to as an aggregation level and must belong to the set $\{1,2,4,8\}$. Let $\Dc$ denote the set of all possible such PDCCHs. For each user the BS first decides an aggregation level, based on its average (long-term) SINR. Then, using that users' unique identifier (ID)  together with its aggregation level, the BS obtains a small subset of non-overlapping PDCCHs from $\Dc$ (of cardinality no greater than $6$) that are eligible to be assigned to that user. Let $\Dc_u$ denote this subset of eligible PDCCHs for a user $u$. Then, if user $u$ is scheduled only one PDCCH from $\Dc_u$ must be assigned to it, i.e., must be used to convey its UL grant. Note that while the PDCCHs that belong to the eligible set of any one user are non-overlapping, those that belong to eligible sets of any two different users can overlap.
 As a result, the BS scheduler must also enforce the constraint that two PDCCHs that are assigned to two different scheduled users, respectively,  must not overlap.

  Next, the constraint that each scheduled user can be assigned only one PDCCH from its set of eligible PDCCHs can be enforced as follows. First, define a set $\Vc_u$ containing $|\Dc_u|$ {\em virtual users} for each user $u,\;1\leq u\leq K$, where each virtual user in $\Vc_u$ is associated with a unique PDCCH in $\Dc_u$ and all the parameters (such as uplink channels, queue size etc.)  corresponding to each virtual user in $\Vc_u$ are identical to those of user $u$.
Let $\tilde{\Uk}$ be the set of all possible subsets of such virtual users, such that each subset has a cardinality no greater than $T$ and contains no more than one virtual user corresponding to the same user. Defining $\tilde{\Mk}=\tilde{\Uk}\times\Ck$,  we can then pose (P1) over $\tilde{\Mk}$ after setting $L=K$ with $\Gc_s=\Vc_s,\;1\leq s\leq K$. Consequently, by defining the virtual users corresponding to each user as being mutually incompatible, we have enforced the constraint that at-most one virtual user for each user can be selected, which in turn is equivalent to enforcing that each scheduled user can be assigned only one PDCCH from its set of eligible PDCCHs.

Finally, consider the set of all eligible PDCCHs, $\{\Dc_u\}_{u=1}^K$. Note that this set is decided by the set of active users and their long-term SINRs. 
Recall that each PDCCH in $\{\Dc_u\}_{u=1}^K$ maps to a unique virtual user.
To ensure that PDCCHs that are assigned to two virtual users corresponding to two different users do not  overlap, we can define multiple binary knapsack constraints. Clearly $R$ such knapsack constraints suffice (indeed can be much more than needed), where each constraint corresponds to one CCE and    has a weight of one for every pair $(\tilde{\Uc},\cb)\in\tilde{\Mk}$ wherein $\tilde{\Uc}$ contains a virtual user   corresponding to a  PDCCH which includes that CCE. Then, a useful consequence of the fact that in LTE the set $\Dc_u$ for each user $u$ is extracted from $\Dc$ via a well designed hash function (which accepts each user's unique ID as input), is that these resulting knapsack constraints are {\em column-sparse}.

\section{Appendix: Proposition I and its Proof}
\begin{proposition}
Let $\hat{W}^{\rm opt,narrow}$ denote the optimal weighted sum rate obtained by solving (P1) albeit where  all pairs $(\Uc,\cb)$ are restricted to lie in $\Mk^{\rm narrow}$. Then, we have that
 \begin{eqnarray}\label{eq:pRAlte2}
  \hat{W}^{\rm narrow}\geq
  \frac{\hat{W}^{\rm opt,narrow}}{1+T+\Delta+2J}.
 \end{eqnarray}
\end{proposition}
\proof Note that Algorithm IIa builds up the stack $\Sc$ in $N$ steps. In particular let $S_j,\;j=1,\cdots,N$ be the element that is added in the $j^{th}$ step and note that either $S_j=\phi$ or it is equal to some pair $(\Uc_j^*,\cb_j^*)$. We use two functions $p^{(j)}_1: \Mk^{\rm narrow}\to\Reals_+$ and $p^{(j)}_2: \Mk^{\rm narrow}\to \Reals_+$ for $j=0,\cdots,N$ to track the function $p'(,)$ as the stack $\Sc$ is being built up over $N$ steps and in particular we set
 $p^{(0)}_1(\Uc,\cb)=0,\;\forall\;(\Uc,\cb)\in\Mc^{\rm narrow}$ and
  $p^{(0)}_2(\Uc,\cb)=p(\Uc,\cb),\;\forall\;(\Uc,\cb)\in\Mc^{\rm narrow}$. 
   For our problem at hand, we define $\{p^{(j)}_1(\Uc,\cb),p^{(j)}_2(\Uc,\cb)\}$ recursively as
  \begin{eqnarray}\label{eq:pRAlte3}
  \nonumber p^{(j)}_1(\Uc,\cb) &=& \left\{\begin{array}{c} (p^{(j-1)}_2(\Uc_j^*,\cb_j^*))^+\Xc\left(p^{(j-1)}_2(\Uc,\cb)>0\right),\;\; \;{\rm If}\;\cb_j^*\cap\cb\neq \phi \\
  (p^{(j-1)}_2(\Uc_j^*,\cb_j^*))^+\Xc\left(p^{(j-1)}_2(\Uc,\cb)>0\right),
  {\rm Else If}\;\exists\;\Gc_s:\Uc\cap\Gc_s\neq\phi\;\&\;\Uc^*_j\cap\Gc_s\neq \phi\\
  (p^{(j-1)}_2(\Uc_j^*,\cb_j^*))^+\Xc\left(p^{(j-1)}_2(\Uc,\cb)>0\right),\;\; {\rm Else If}\;\exists\;q\in\Ic:\alpha^q( \Uc,\cb)=\alpha^q(\Uc_j^*,\cb_j^*)=1\\
   2(p^{(j-1)}_2(\Uc_j^*,\cb_j^*))^+\Xc\left(p^{(j-1)}_2(\Uc,\cb)>0\right)\max_{1\leq q\leq J}\beta^q( \Uc,\cb),\;\;{\rm Otherwise}\end{array}\right.\\
   p^{(j)}_2(\Uc,\cb)&=& p^{(j-1)}_2(\Uc,\cb) - p^{(j)}_1(\Uc,\cb),
 \end{eqnarray}
    where $(x)^+=\max\{x,0\},\;x\in\Reals$, $\Xc(.)$ denotes the indicator function and $(\Uc_j^*,\cb_j^*)=\arg\max_{(\Uc,\cb)\in\Mk^{\rm narrow}\atop {\rm Tail}(\cb)=j}p^{(j-1)}_2 (\Uc,\cb)$. Hence, we have that
    \begin{eqnarray}
  p^{(j-1)}_2(\Uc,\cb) =  p^{(j)}_2(\Uc,\cb) + p^{(j)}_1(\Uc,\cb),\;\;\forall\; (\Uc,\cb)\in\Mk^{\rm narrow},\;j=1,\cdots,N.
 \end{eqnarray}
 It can be noted that
  \begin{eqnarray}\label{eq:pRAlte4}
 \nonumber p^{(j)}_2(\Uc,\cb) \leq 0,\;\;\forall\; (\Uc,\cb)\in\Mk^{\rm narrow}: {\rm Tail}(\cb)\leq j\\
  p^{(k)}_2(\Uc,\cb) \leq p^{(j)}_2(\Uc,\cb),\;\;\forall\; (\Uc,\cb)\in\Mk^{\rm narrow}\;\&\;k\geq j.
 \end{eqnarray}
   Further, to track the stack $\Sc'$ which is built in the while loop of the algorithm, we define stacks $\{\Sc^*_j\}_{j=0}^N$ where $\Sc_N^*=\phi$ and $\Sc_j^*$ is the value of $\Sc'$ after the Algorithm has tried to add $\cup_{m=j+1}^N S_m$ to $\Sc'$ (starting from $\Sc'=\phi$) so that $\Sc_0^*$ is the stack $\Sc'$ that is the output of the Algorithm. Note that  $\Sc_{j+1}^*\subseteq\Sc_j^*\subseteq \Sc_{j+1}^*\cup S_{j+1}$.
   Next, for $j=0,\cdots,N$, we let $W^{(j)\;{\rm opt}}$ denote the optimal solution to (P1) but where $\Mk$ is replaced by
    $\Mk^{\rm narrow}$ and the function $p(,)$ is replaced by $p_2^{(j)}(,)$. Further, let $W^{(j)}=\sum_{(\Uc,\cb)\in \Sc_j^*}p_2^{(j)}(\Uc,\cb)$ and note that $\hat{W}^{\rm opt, narrow}=W^{(0)\;{\rm opt}}$ and $\hat{W}^{\rm  narrow}=W^{(0)}$.
    We will show via induction that
 \begin{eqnarray}\label{eq:pRAlte5}
 W^{(j)\;{\rm opt}}\leq (T+1+\Delta+2J)W^{(j)},\;\forall\;j=N,\cdots,0,
  \end{eqnarray}
  which includes the claim in (\ref{eq:pRAlte2}) at $j=0$.
   First note that  the base case $W^{(N)\;{\rm opt}}\leq (T+1+\Delta+2J)W^{(N)}$ is readily true since $\Sc_N^*=\phi$ and
      $p^{(N)}_2(\Uc,\cb) \leq 0,\;\;\forall\; (\Uc,\cb)\in\Mk^{\rm narrow}$. Then,
      assume that (\ref{eq:pRAlte5}) holds for some $j$. We focus only on the main case in which $S_j=(\Uc_j^*,\cb_j^*)\neq \phi$ (the remaining case holds trivially true). Note that
        since $(\Uc_j^*,\cb_j^*)$ is added to the stack $\Sc$ in the algorithm, $p^{(j-1)}_2 (\Uc_j^*,\cb_j^*)>0$. Then from the update formulas (\ref{eq:pRAlte3}), we must have
        that  $p^{(j)}_2 (\Uc_j^*,\cb_j^*)=0$. Using the fact that  $\Sc_{j-1}^*\subseteq \Sc_{j}^*\cup (\Uc_j^*,\cb_j^*)$ together with the induction hypothesis, we can conclude that
        \begin{eqnarray}\label{eq:pRAlte5b}
   W^{(j)}=\sum_{(\Uc,\cb)\in \Sc_j^*}p_2^{(j)}(\Uc,\cb)=\sum_{(\Uc,\cb)\in \Sc_{j-1}^*}p_2^{(j)}(\Uc,\cb)\geq
    \frac{W^{(j)\;{\rm opt}}}{T+1+\Delta+2J}.
   \end{eqnarray}

Upon  invoking Lemma \ref{lem:prophelp}, which is stated and proved below,   we obtain that
        \begin{eqnarray}\label{eq:pRAlte6}
   \sum_{(\Uc,\cb)\in \Sc_{j-1}^*}p_1^{(j)}(\Uc,\cb)\geq p^{(j-1)}_2 (\Uc_j^*,\cb_j^*).
   \end{eqnarray}
     Then, let  $V^{(j)\;{\rm opt}}$ denote the optimal solution to (P1) but where $\Mk$ is replaced by
    $\Mk^{\rm narrow}$ and the function $p(,)$ is replaced by $p_1^{(j)}(,)$. Upon invoking Lemma \ref{lem:prophelp2}, also stated and proved below,  we can conclude that
     \begin{eqnarray}\label{eq:pRAlte7}
    p^{(j-1)}_2 (\Uc_j^*,\cb_j^*)\geq
    \frac{V^{(j)\;{\rm opt}}}{T+1+\Delta+2J}.
   \end{eqnarray}

Thus, using (\ref{eq:pRAlte5b}), (\ref{eq:pRAlte6}) and (\ref{eq:pRAlte7}) we can conclude that
\begin{eqnarray}
    (1+T+\Delta+2J) \sum_{(\Uc,\cb)\in \Sc_{j-1}^*}(\underbrace{p_1^{(j)}(\Uc,\cb)+p_2^{(j)}(\Uc,\cb)}_{p_2^{(j-1)}(\Uc,\cb)}\geq  V^{(j)\;{\rm opt}} + W^{(j)\;{\rm opt}}\geq   W^{(j-1)\;{\rm opt}}.
   \end{eqnarray}
which proves the induction step and proves the claim in (\ref{eq:pRAlte2}). \endproof

\begin{lemma}\label{lem:prophelp}
For all $j$ we have that
  \begin{eqnarray}\label{eq:pRAlte6l}
   \sum_{(\Uc,\cb)\in \Sc_{j-1}^*}p_1^{(j)}(\Uc,\cb)\geq p^{(j-1)}_2 (\Uc_j^*,\cb_j^*).
   \end{eqnarray}
\end{lemma}
 \proof
     Suppose that $\Sc_{j-1}^*= \Sc_{j}^*\cup (\Uc_j^*,\cb_j^*) $. Then, recalling  (\ref{eq:pRAlte3}) we can deduce that  (\ref{eq:pRAlte6l}) is true since $p_1^{(j)}(\Uc_j^*,\cb_j^*)= p^{(j-1)}_2 (\Uc_j^*,\cb_j^*)$.
        Suppose now that $\Sc_{j-1}^*= \Sc_{j}^*$. In this case we can have two possibilities. In the first one $(\Uc_j^*,\cb_j^*)$ cannot not be added to $\Sc_{j}^*$ due to the presence of a pair
         $(\Uc',\cb')\in\Sc_j^*$ for which at-least one of these three conditions are satisfied: $\exists\;\Gc_s:\Uc'\cap\Gc_s\neq\phi\;\&\;\Uc^*_j\cap\Gc_s\neq \phi$; $\cb'\cap\cb_j^*\neq \phi$ and $\exists\;q\in\Ic:\alpha^q( \Uc',\cb')=\alpha^q(\Uc_j^*,\cb_j^*)$=1 . Since any pair $(\Uc',\cb')\in\Sc_j^*$ was added to $\Sc$ in the algorithm after the $j^{th}$ step, from the second inequality in (\ref{eq:pRAlte4}) we must have that $p^{(j-1)}_2(\Uc',\cb')>0$. Recalling (\ref{eq:pRAlte3}) we can then deduce that $p_1^{(j)}(\Uc',\cb')= p^{(j-1)}_2 (\Uc_j^*,\cb_j^*)$ which proves (\ref{eq:pRAlte6l}). In the second possibility, $(\Uc_j^*,\cb_j^*)$ cannot not be added to $\Sc_{j}^*$ due to a generic knapsack constraint being violated. In other words, for some $q\in\{1,\cdots,J\}$, we have that
         \begin{eqnarray}
    \sum_{(\Uc ,\cb)\in\Sc_j^*}\beta^q(\Uc ,\cb)>1-\beta^q(\Uc_j^*,\cb_j^*).
   \end{eqnarray}
           Since $(\Uc_j^*,\cb_j^*)\in\Mk^{\rm narrow}$, $\beta^q(\Uc_j^*,\cb_j^*)\leq 1/2$ so that
        \begin{eqnarray}
   2\sum_{(\Uc ,\cb)\in\Sc_j^*}\max_{1\leq q\leq J}\beta^q(\Uc ,\cb)\geq 2 \sum_{(\Uc ,\cb)\in\Sc_j^*}\beta^q(\Uc ,\cb)>1,
   \end{eqnarray}
which along with (\ref{eq:pRAlte3}) also proves (\ref{eq:pRAlte6l}). Thus, we have established the claim in (\ref{eq:pRAlte6l}).\endproof
\begin{lemma}\label{lem:prophelp2}
Let  $V^{(j)\;{\rm opt}}$ denote the optimal solution to (P1) but where $\Mk$ is replaced by
    $\Mk^{\rm narrow}$ and the function $p(,)$ is replaced by $p_1^{(j)}(,)$. Then,  we have that
     \begin{eqnarray}\label{eq:pRAlte7l}
    p^{(j-1)}_2 (\Uc_j^*,\cb_j^*)\geq
    \frac{V^{(j)\;{\rm opt}}}{T+1+\Delta+2J}.
   \end{eqnarray}
\end{lemma}
\proof
 First, from (\ref{eq:pRAlte3}) we note that for any pair $(\Uc,\cb)\in\Mk^{\rm narrow}$, $p_1^{(j)}(\Uc,\cb)\leq p^{(j-1)}_2 (\Uc_j^*,\cb_j^*)$. Let $\Vc_1^{(j)\;{\rm opt}}$ be an optimal allocation of pairs that results in $V^{(j)\;{\rm opt}}$. For any two pairs $(\Uc_1,\cb_1),(\Uc_2,\cb_2)\in \Vc_1^{(j)\;{\rm opt}}$ we must have that for each $\Gc_s\;1\leq s\leq L$, at-least one of $\Uc_1\cap\Gc_s$ and $\Uc_2\cap\Gc_s$ is  $\phi$,  as well as $\cb_1\cap\cb_2=\phi$. In addition, $|\Uc_1|$ and $|\Uc_2|$ are no greater than $T$. Thus we can have at-most $T$ such pairs $\{(\Uc,\cb)\}$ in $\Vc_1^{(j)\;{\rm opt}}$ for which  $\exists\;\Gc_s:\Uc\cap\Gc_s\neq\phi\;\&\;\Uc^*_j\cap\Gc_s\neq \phi$. Further, using the first inequality in (\ref{eq:pRAlte4}) we see that any pair $(\Uc,\cb)$ for which $\cb\cap\cb_j^*\neq \phi$ and $p_1^{(j)}(\Uc,\cb)= p^{(j-1)}_2 (\Uc_j^*,\cb_j^*)$ must have ${\rm Tail}(\cb)\geq j$ so that $j\in\cb$. Thus, $\Vc_1^{(j)\;{\rm opt}}$ can include at-most one pair $(\Uc,\cb)$ for which $\cb\cap\cb_j^*\neq \phi$. Next, there can be at-most $\Delta$ constraints in $\Ic$ for which $\alpha^q(\Uc_j^*,\cb_j^*)=1,q\in\Ic$ is satisfied. For each such constraint $q\in\Ic$ we can pick at-most one pair $(\Uc,\cb)$ for which $\alpha^q(\Uc,\cb)=1$ and $p_1^{(j)}(\Uc,\cb)= p^{(j-1)}_2 (\Uc_j^*,\cb_j^*)$. Thus, $\Vc_1^{(j)\;{\rm opt}}$ can include at-most $\Delta$ such pairs, one for each constraint.
  Now the remaining pairs in $\Vc_1^{(j)\;{\rm opt}}$ (whose users do not intersect any group $\Gc_s\;1\leq s\leq L$ that $\Uc_j^*$ does and whose chunks do not intersect $\cb_j^*$ and which do not violate any binary knapsack constraint in the presence of $(\Uc_j^*,\cb_j^*)$) must satisfy the generic knapsack constraints. Let these pairs form the
 set $\tilde{\Vc}_1^{(j)\;{\rm opt}}$ so that,
 \begin{eqnarray*}
     \sum_{(\Uc,\cb)\in \tilde{\Vc}_1^{(j)\;{\rm opt}}}p_1^{(j)}(\Uc,\cb)=\sum_{(\Uc,\cb)\in \tilde{\Vc}_1^{(j)\;{\rm opt}}}2p^{(j-1)}_2 (\Uc_j^*,\cb_j^*)\max_{1\leq q\leq J}\beta^q(\Uc ,\cb)\leq 2p^{(j-1)}_2 (\Uc_j^*,\cb_j^*)\sum_{q=1}^J\sum_{(\Uc,\cb)\in \tilde{\Vc}_1^{(j)\;{\rm opt}}}\beta^q(\Uc ,\cb)\\
     \leq 2Jp^{(j-1)}_2 (\Uc_j^*,\cb_j^*).
   \end{eqnarray*}
Combining these observations we have that
\begin{eqnarray}
  V^{(j)\;{\rm opt}}=   \sum_{(\Uc,\cb)\in \Vc_1^{(j)\;{\rm opt}}}p_1^{(j)}(\Uc,\cb)\leq  (1+T+\Delta+2J)p^{(j-1)}_2 (\Uc_j^*,\cb_j^*),
   \end{eqnarray}
which is the desired result in (\ref{eq:pRAlte7l}).\endproof

\end{document}